\documentclass[prb,twocolumn,superscriptaddress]{revtex4}

\usepackage{amsmath}
\usepackage{amsfonts}
\usepackage{amssymb,mathrsfs}
\usepackage{bbold,yfonts}
\usepackage{upgreek}
\usepackage{graphicx}
\usepackage{xcolor}

\newcommand{\bs}[1]{\boldsymbol{#1}}

\begin{document}

\title{Optical conductivity in graphene: hydrodynamic regime}

\author{B.N. Narozhny}
\affiliation{\mbox{Institut f\"ur Theorie der kondensierten Materie, Karlsruhe Institute of
Technology, 76128 Karlsruhe, Germany}}
\affiliation{National Research Nuclear University MEPhI (Moscow Engineering Physics Institute),
  115409 Moscow, Russia}

\date{\today}

\begin{abstract}
   A recent measurement of the optical conductivity in graphene
   [P. Gallagher {\it et.al}, Science {\bf 364}, 158 (2019)] offers a
   possibility of experimental determination of microscopic time
   scales describing scattering processes in the electronic fluid. In
   this paper, I report a theoretical calculation of the optical
   conductivity in graphene at arbitrary doping levels, within the
   whole ``hydrodynamic'' temperature range, and for arbitrary
   non-quantizing magnetic fields. The obtained results are in good
   agreement with the available experimental data.
\end{abstract}

\maketitle

Recent experiments \cite{gal,geim1,kim1,kim2,geim2,geim3,sulp,geim4}
indicate that charge carriers in graphene at nearly room temperatures
may exhibit a hydrodynamic flow \cite{rev,luc}. Traditional
linear-response transport measurements uncovered such remarkable
features as the strong violation of the Wiedemann-Franz law
\cite{kim1} and superballistic transport \cite{geim2}. In this type of
experiments all information about the flow of electrons is extracted
from a small number of measured resistances
\cite{geim1,kim2,geim2,geim4} and Johnson noise power
\cite{kim1}. Additional information about the flow can be obtained by
current density imaging \cite{sulp,imh,imm} or terahertz spectroscopy
\cite{gal}. The latter experiment yields the optical conductivity
which can be used to extract information about the electron-electron
and electron-impurity scattering rates in graphene \cite{gal}.

Hydrodynamic theory of electronic transport \cite{rev,luc} can be
formulated similarly to the usual hydrodynamics \cite{dau6} with the
important caveat: the total momentum of the electronic system in a
solid is not, strictly speaking, a conserved quantity. In general,
this means that one can only hope to observe electronic hydrodynamics
in an intermediate temperature window where the electron-electron
interaction is the dominant scattering mechanism in the problem
\cite{rev}. In the specific case of graphene, there is an additional
circumstance due to the linearity of the excitation spectrum: the
momentum density is proportional to the energy current
\cite{rev,luc,hydro1,me1,me2} (rather than the ``mass'' current as in
the usual hydrodynamics \cite{dau6}). Hence it is the electrical
(rather than the thermal) conductivity that appears in the
hydrodynamic theory of electronic transport in graphene as a
dissipative coefficient \cite{hydro1,me1,me2}.

In traditional hydrodynamics, dissipative processes are described by
three coefficients \cite{dau6}: the shear and bulk viscosities and the
thermal conductivity. In graphene, the bulk viscosity vanishes
\cite{rev,luc,me1}, leaving the shear viscosity as the only
dissipative coefficient in the generalized Navier-Stokes equation
\cite{hydro1,me1,me2}. Under the assumption of approximate
conservation of the particle number in each of the two bands in
graphene \cite{alf}, electronic hydrodynamics describes two
macroscopic quasiparticle currents (the electric and ``imbalance''
currents \cite{hydro1,me1,me2,alf}). Dissipative corrections to these
currents due to electron-electron interaction are described by a
${2\times2}$ matrix of coefficients \cite{luc}. In addition, disorder
scattering not only contributes to these coefficients, but also
determines a correction to the energy current \cite{me1} yielding the
thermal conductivity \cite{alf}. As a result, the electric and energy
currents are relaxed by two different scattering mechanisms leading to
the violation of the Wiedemann-Franz law \cite{kim1}. The effect is
especially pronounced at charge neutrality \cite{kim1} where the two
currents are completely decoupled.

Assuming the applicability of the kinetic approach, one can derive the
dissipative coefficients appearing in hydrodynamics starting with the
Boltzmann equation and the local equilibrium distribution function
\cite{rev,luc,hydro1,me1,me2,dau10}. The resulting viscosities and
conductivities are temperature- and material-dependent constants
\cite{dau10}. In contrast, calculations based on the Kubo formula
yield frequency-dependent viscosities \cite{read,poli16,julia1} and
conductivities \cite{read,Giuliani}. The frequency-dependent (optical)
conductivity is experimentally measurable \cite{Giuliani} even in the
hydrodynamic regime \cite{gal}.

In this paper I extend the kinetic derivation of the dissipative
corrections to the hydrodynamic quantities in graphene allowing for
the low-frequency optical conductivity (in general, hydrodynamics is
valid for frequencies which are much lower than the typical scattering
rate associated with equilibration processes,
${\omega\tau_{ee}\ll1}$). I also use the kinetic theory to extend the
hydrodynamic-like macroscopic description \cite{hydro0} to higher
frequencies where the results should be compared to those in the
high-frequency collisionless regime
\cite{mish,teber,ssh,julia2,geimop,geimop1,julia1}.

Previously, optical conductivity in the hydrodynamic regime in
graphene was addressed in Refs.~\onlinecite{har,mfs,mfss,mus,fog}.  A
large part of that research was focused on the two limiting cases,
charge neutrality and the degenerate regime. In the absence of the
magnetic field, the results for these two limits can be combined in an
interpolation formula \cite{fog}
\begin{equation}
\label{s0}
\sigma(\omega; \bs{q}\!=\!0) = \frac{A_1}{-i\omega\!+\!\tau_{\rm dis}^{-1}}
+
\frac{A_2}{-i\omega\!+\!\tau_{ee}^{-1}\!+\!\tau_{\rm dis}^{-1}},
\end{equation}
where $A_i$ are temperature- and density-dependent constants,
$\tau_{\rm dis}$ is the elastic mean free time, and $\tau_{ee}$ is the
typical time scale associated with the electron-electron interaction.
Here I report the results of a rigorous calculation of
$\sigma(\omega)$ for arbitrary carrier densities, temperatures (within
the hydrodynamic range), and classical (non-quantizing) magnetic
fields beyond the limiting behavior of Eq.~(\ref{s0}), as well as of
the microscopic times scales determining $\sigma(\omega)$.

\section{Unconventional hydrodynamics in graphene}

Unconventional hydrodynamics in graphene was derived in
Refs.~\onlinecite{hydro1,me1} on the basis of the kinetic theory.  The
need for the derivation was dictated by the lack of symmetry in the
problem -- the electronic system in graphene is neither Galilean, nor
Lorentz invariant. The former follows from the linearity of the
excitation spectrum, while the latter is related to the classical,
three-dimensional nature of the Coulomb interaction in graphene.

As in any other solid, electrons in graphene may scatter on lattice
vibrations (phonons) and imperfections (typically referred to as
``disorder'') and hence loose momentum. In this case, one can speak of
hydrodynamics only within a limited parameter regime (e.g., at
intermediate temperatures) where the electron-electron interaction is
the dominant scattering mechanism in the problem. This
``hydrodynamic'' regime can be defined by the inequality
\begin{equation}
\label{con1}
\tau_{\rm ee} \ll \tau_{\rm dis}, \tau_{\rm e-ph}, \tau_R, \,\, {\rm etc},
\end{equation}
where $\tau_{\rm e-ph}$ is the typical scale describing the
electron-phonon interaction, $\tau_R$ is the quasiparticle
``recombination'' time and ``etc'' stands for any other
scattering-related time scale in the problem.

Assuming the applicability of the kinetic (Boltzmann) equation at
least in some subset of the hydrodynamic region, hydrodynamic
equations can be derived from the kinetic equation following the
standard procedure \cite{dau10}.

Conservation of charge and energy can be expressed in terms of the
continuity equations \cite{hydro1,me1,me2}
\begin{subequations}
\label{ces}
\begin{equation}
\label{cen1}
\partial_t n + \bs{\nabla}_{\bs{r}}\!\cdot\!\bs{j} = 0,
\end{equation}
\begin{equation}
\label{cene1}
\partial_t n_E + \bs{\nabla}_{\bs{r}}\!\cdot\!\bs{j}_E = e \bs{E}\cdot\bs{j}.
\end{equation}
where $n$ is the carrier density, $\bs{j}$ is the current (differing
from the charge density and electric current by a multiplicative
factor of the electric charge, $e$), $n_E$ and $\bs{j}_E$ are the
energy density and current, and $\bs{E}$ is the electric field
($e\bs{E}\cdot\bs{j}$ describing Joule heat). The continuity equation
(\ref{cen1}) is valid in any electronic system, while
Eq.~(\ref{cene1}) neglects possible energy losses due to coupling to
collective excitations (e.g., phonons, plasmons, etc.).

The main equations of the hydrodynamic theory, the Euler and
Navier-Stokes equations (for ideal and viscous fluids, respectively),
are based on the continuity equation for momentum density reflecting
momentum conservation. In solids, electronic momentum can be treated
as a conserved quantity only approximately, in the sense of
Eq.~(\ref{con1}). Hence, the generalized Navier-Stokes equation
in graphene \cite{me1} contains a weak disorder scattering term,
\begin{eqnarray}
\label{eq1g}
&&
\!\!\!\!\!\!
W(\partial_t+\bs{u}\!\cdot\!\bs{\nabla})\bs{u}
+
v_g^2 \bs{\nabla} P
+
\bs{u} \partial_t P 
+
e(\bs{E}\!\cdot\!\bs{j})\bs{u} 
=
\\
&&
\nonumber\\
&&
=
v_g^2 
\left[
\eta \Delta\bs{u}
-
\eta_H \Delta\bs{u}\!\times\!\bs{e}_B
+
en\bs{E}
+
\frac{e}{c} \bs{j}\!\times\!\bs{B}
\right]
-
\frac{\bs{j}_E}{\tau_{{\rm dis}}},
\nonumber
\end{eqnarray}
where $v_g$ is the Fermi velocity in graphene, $\bs{u}$ is the
hydrodynamic velocity, $\bs{B}$ is the magnetic field, $c$ is the
speed of light, and $W$ and $P$ are the enthalpy and pressure, which
are related to the energy density in graphene by the ``equation of
state'' \cite{hydro1,me1}
\begin{equation}
\label{eqsta}
W=n_{E}+P = \frac{3 n_{E}}{2\!+\!u^2/v_g^2}.
\end{equation}
The shear, $\eta$, and Hall, $\eta_H$, viscosities in graphene were
discussed theoretically in Ref.~\onlinecite{me2} and experimentally in
Refs.~\onlinecite{geim1,geim4}.

Due to kinematic suppression of interband scattering in graphene
\cite{alf,luc}, one may consider conservation of the particle number
in each band. This leads to another continuity equation
\begin{equation}
\label{ceni1}
\partial_t n_I + \bs{\nabla}_{\bs{r}}\!\cdot\!\bs{j}_I = -
\frac{n_I\!-\!n_{I,0}}{\tau_R},
\end{equation}
where $n_{I,0}$ is the equilibrium ``imbalance'' density. The
``imbalance'' density and current \cite{alf} (as well as $n$ and
$\bs{j}$) are related to the densities and currents in each band as
\begin{equation}
\label{ds}
n=n_+-n_-, \qquad n_I = n_++n_-,
\end{equation}
\begin{equation}
\bs{j} = \bs{j}_+ - \bs{j}_-,
\qquad
\bs{j}_I = \bs{j}_+ + \bs{j}_-.
\end{equation}

Finally, the quasiparticle currents $\bs{j}$ and $\bs{j}_I$ are not
related to any conserved quantity and hence acquire dissipative
corrections (similarly to the energy current in the traditional
hydrodynamics \cite{dau6}). The energy current in graphene is
proportional to the momentum density and hence cannot be relaxed by
electron-electron interaction. However, it can be relaxed by weak
disorder (leading to the strong violation of the Wiedemann-Franz law
\cite{kim1}). Combining the dissipative corrections to the three
macroscopic currents in graphene, one defines the second set of
dissipative coefficients \cite{me1}
\begin{equation}
\label{djdef}
\bs{j} = n \bs{u} + \delta\bs{j},
\quad
\bs{j}_I = n_{I} \bs{u} + \delta\bs{j}_I,
\quad
\bs{j}_{E} = W\bs{u} + \delta\bs{j}_E,
\end{equation}
\begin{equation}
\label{djs}
\begin{pmatrix}
  \delta\bs{j} \cr
  \delta\bs{j}_I \cr
  \delta\bs{j}_E/T
\end{pmatrix}
=
\widehat\Sigma\!
\begin{pmatrix}
\bs{F}_{\!\bs{u}}\!-\!T\,\bs{\nabla}\frac{\mu}{T} \cr
T\,\bs{\nabla}\frac{\mu_I}{T} \cr
0
\end{pmatrix}
+
\widehat\Sigma_H\!
\begin{pmatrix}
\bs{F}_{\!\bs{u}}\!-\!T\,\bs{\nabla}\frac{\mu}{T} \cr
T\,\bs{\nabla}\frac{\mu_I}{T} \cr
0
\end{pmatrix}\!\times\bs{e}_B,
\end{equation}
where
${\bs{F}_{\!\bs{u}}=e\bs{E}\!+\!\frac{e}{c}\bs{u}\!\times\!\bs{B}}$,
while 
$\mu$ and $\mu_I$ are the chemical potentials conjugate to
$n$ and $n_I$, respectively. They are related to the
chemical potentials $\mu_\lambda$ as
\begin{equation}
\label{mu}
\mu=(\mu_++\mu_-)/2, \qquad
\mu_I=(\mu_+-\mu_-)/2.
\end{equation}
\end{subequations}
In this paper I disregard thermoelectric effects \cite{alf} and set
$\mu_\pm=\mu$ (or $\mu_I=0$).

In addition to Eqs.~(\ref{ces}) the complete hydrodynamic theory
includes Maxwell's equations taking into account electromagnetic
fields induced by inhomogeneities of the charge density similarly to
the Vlasov self-consistency \cite{dau10}. In solids typical velocities
are rather small, ${v_g\ll{c}}$, and hence the Maxwell's equations are
usually reduced to electrostatics. In gated structures
\cite{geim1,geim2,geim3,geim4} the electrostatics is controlled by
the gate \cite{ash,mr1} further simplifying the relation between the
charge density and electric field.

\section{Dissipative coefficients}

The dissipative coefficients $\widehat\Sigma$ and $\widehat\Sigma_H$
are related to the electrical conductivity \cite{luc} (and are the
counterpart of the thermal conductivity in traditional hydrodynamics
\cite{dau6}). In the absence of the magnetic field the Hall term
vanishes, ${\widehat\Sigma_H(B\!=\!0)=0}$. The matrix
${\widehat\Sigma(B\!=\!0)}$ simplifies at the Dirac point, where it is
block diagonal, such that the electric current decouples from the
other two. Moreover, at ${n=0}$ the ``hydrodynamic'' contribution to
$\bs{j}$ vanishes [see Eq.~(\ref{djdef})], leaving the dissipative
correction $\delta\bs{j}$ as the total current. This correction
remains finite even in the absence if disorder,
\begin{equation}
\label{sq}
e\delta\bs{j} (\mu\!=\!0) = \sigma_Q\bs{E},
\quad
\sigma_Q = {\cal A}e^2/\alpha_g^2,
\quad
{\cal A}\approx0.12,
\end{equation}
where $\sigma_Q$ is known as the ``quantum'' or ``intrinsic''
conductivity of graphene
\cite{kash,har,mfs,mfss,mus,schutt,hydro1,luc,me1}. The quantity
$\sigma_Q$ is a constant that depends on temperature only through the
logarithmic renormalization of the coupling constant $\alpha_g$
\cite{shsch}. Away from the Dirac point, matrix elements of
${\widehat\Sigma}$ may exhibit more pronounced temperature and density
dependence, but within the framework of Ref.~\onlinecite{me1} they
remain independent of frequency.

\subsection{Kinetic theory approach}

Within the standard approach \cite{dau10}, one derives the
hydrodynamic theory from the kinetic equation
\begin{equation}
\label{ke0}
{\cal L}f \!=\! {\rm St} [f],
\quad
{\cal L} \!=\! \partial_t \!+\! \bs{v}\!\cdot\!\bs{\nabla}_{\bs{r}} \!+\! 
(e\bs{E} \!+\! \frac{e}{c} \bs{v}\!\times\!\bs{B})\!\cdot\!\bs{\nabla}_{\bs{k}}.
\end{equation}
Under the assumption of local equilibrium, Eq.~(\ref{ke0}) can be
solved approximately
\begin{equation}
\label{df0}
f = f^{(0)} + \delta f,
\end{equation}
where $f^{(0)}$ is the local equilibrium distribution function
(nullifying the collision integral due to electron-electron
interaction) and $\delta f$ is the nonequilibrium correction. The
latter can be found within linear response:
\begin{equation}
\label{ke01}
{\cal L}f^{(0)} = {\rm St} [\delta f].
\end{equation}
Linearizing the collision integral, one proceeds with the solution of
the resulting linear integral equation (which, however nontrivial, is
much simpler than the original nonlinear integro-differential
equation). Macroscopic description corresponding to local equilibrium
is the ideal (Euler) hydrodynamics, while the nonequilibrium
correction is responsible for dissipative terms.

The reason one is justified in neglecting ${\cal L}\delta f$ in
Eq.~(\ref{ke01}) is the long wavelength nature of hydrodynamics --
macroscopic quantities are assumed to be varying slowly over long
distances such that their gradients are small. Hence, gradients of the
correction, $\bs{\nabla}\delta f$, represent the second-order
smallness in the hydrodynamic expansion (formally, in the so-called
Knudsen number \cite{dau6,dau10}). Similar argument can be applied to
the electric field. Magnetic field is typically not treated within
linear response \cite{me1} leading to the field-dependent viscosity
and Hall viscosity \cite{me2,geim4,pol17,ady,ady2}. Treating the time
derivative in the Liouville's operator in the same manner leads to
frequency-dependent dissipative coefficients. In other words, instead
of Eq.~(\ref{ke01}) yielding constant, field-independent viscosity and
conductivity, one should solve the equation
\begin{equation}
\label{ke02}
{\cal L}\Big|_{\bs{B}=0}f^{(0)} \!+ \partial_t \delta f
\!+ \frac{e}{c}\left[\bs{v}\!\times\!\bs{B}\right]\cdot\bs{\nabla}_{\bs{k}} f
= {\rm St}_{ee} [\delta f] + {\rm St}_{\rm dis} [f].
\end{equation}
In this paper, I solve Eq.~(\ref{ke02}) focusing on the frequency- and
field-dependent ``effective conductivities'' $\widehat\Sigma(\omega)$
and $\widehat\Sigma_H(\omega)$. A similar analysis of the
frequency-dependent viscosity \cite{pal} will be reported elsewhere.

\subsection{Dissipative corrections to macroscopic currents in graphene}

A comprehensive derivation of the dissipative hydrodynamics in
graphene was reported in Ref.~\onlinecite{me1}. In comparison to
Eq.~(\ref{ke02}) that calculation disregarded the time derivative,
$\partial_t\delta f$, yielding frequency-independent dissipative
coefficients. On the other hand, the other terms in the (linear)
equation (\ref{ke02}) were evaluated in all the details. In this
Section I outline that calculation in order to make the present present
paper self-contained focusing on the essential changes that are
necessary to evaluate optical conductivity. Technical details can be
found in Ref.~\onlinecite{me1}.


The main idea of the calculation \cite{hydro1,me1} is to formulate
macroscopic equations for the three currents (\ref{djs}) integrating
the kinetic equation and expressing the non-equilibrium correction to
the distribution function, $\delta f$, in terms of the dissipative
corrections to the currents. To the resulting ``three-mode''
approximation, $\delta f$ is (in graphene, single-particle states can
be labeled by the band index ${\lambda=\pm}$ and momentum $\bs{k}$; in
what follows, these indices are often omitted for brevity).
\begin{subequations}
\label{df}
\begin{equation}
\delta f \!=\! 
f^{(0)} \!\left[1\!-\!f^{(0)}\right]\! h,
\qquad
h_{\lambda\bs{k}} \!=\! \frac{\bs{v}_{\lambda\bs{k}}}{v_g}
\sum_{i=1}^3\phi_i \bs{h}^{(i)},
\end{equation}
with the ``three modes'' (not exhibiting the collinear scattering
singularity \cite{rev,mfss,hydro1,me1}) described by
\[
\phi_1=1, 
\qquad
\phi_2=\lambda,
\qquad
\phi_3=\frac{\epsilon}{T},
\]
where ${\epsilon=\epsilon_{\lambda\bs{k}}}$ is the excitation energy
(${\bs{v}_{\lambda\bs{k}}=\partial\epsilon_{\lambda\bs{k}}/\partial\bs{k}}$)
and the vectors $\bs{h}^{(i)}$ are related to the dissipative
corrections (\ref{djs}) to the macroscopic currents (\ref{djdef}) as
\cite{hydro1,me1}
\begin{equation}
\label{djsm}
\begin{pmatrix}
\delta\bs{j} \cr
\delta\bs{j}_I \cr
\delta\bs{j}_E/T
\end{pmatrix}
=
\frac{v_gT}{2} \widehat{M}_h
\begin{pmatrix}
\bs{h}^{(1)} \cr
\bs{h}^{(2)} \cr
\bs{h}^{(3)}
\end{pmatrix}.
\end{equation}
The matrix $\widehat{M}_h$ is expressed in terms of equilibrium
densities and compressibilities (see Appendix~\ref{leqs} for their
explicit form) as
\begin{equation}
\label{mhdim}
\widehat{M}_h \!=\! \frac{\partial n}{\partial \mu} \widehat{\textswab{M}}_h,
\quad
\widehat{\textswab{M}}_h\!=\!
\begin{pmatrix}
1 & \frac{xT}{\cal T} & 2\tilde{n} \frac{T}{\cal T} \cr
\frac{xT}{\cal T} & 1 & \left[x^2\!+\!\frac{\pi^2}{3}\right]\!\frac{T}{\cal T} \cr
2\tilde{n} \frac{T}{\cal T} & \left[x^2\!+\!\frac{\pi^2}{3}\right]\!\frac{T}{\cal T} &
6\tilde{n}_E \frac{T}{\cal T}
\end{pmatrix}\!,
\end{equation}
\end{subequations}
where $\tilde{n}$ and $\tilde{n}_E$ are the dimensionless carrier and
energy densities, respectively, ${x=\mu/T}$, and
\[
{\cal{T}}=2\pi{v}_g^2\frac{\partial{n}}{\partial\mu}
=
2T\ln\left[2\cosh\frac{\mu}{2T}\right].
\]

Multiplying Eq.~(\ref{ke02}) by the velocity,
$\bs{v}_{\lambda\bs{k}}$, and integrating over all single-particle
states (${N=4}$ is the degeneracy factor), one finds the macroscopic
equation for the electric current
\begin{subequations}
\label{jeq0}
\begin{equation}
\label{jeq01}
N\!\sum_\lambda\!\!\int\!\!\frac{d^2k}{(2\pi)^2} \bs{v}_{\lambda\bs{k}} {\cal L} f^{(0)}_{\lambda\bs{k}}
\!=\!
\bs{\cal I}_1\left[f\right]
\!-\!
\frac{\partial\delta\bs{j}}{\partial t}
\!-\!
\omega_B \bs{e}_B\!\times\!\bs{\cal K}\left[\delta f\right].
\end{equation}
The integrated collision integral, ${\bs{\cal{I}}_1\left[f\right]}$,
comprises the electron-electron and disorder scattering terms 
\begin{equation}
\label{i1def}
\bs{\cal{I}}_1 =
N\!\sum_\lambda\!\!\int\!\!\frac{d^2k}{(2\pi)^2} \bs{v}_{\lambda\bs{k}}
\left({\rm St}_{ee} [\delta f] + {\rm St}_{\rm dis} [f]\right)
\equiv
\bs{\cal I}_1^{ee}+\bs{\cal I}_1^{\rm dis}.
\end{equation}
I assume the latter to be weak [see Eq.~(\ref{con1})] such that the
$\tau$-approximation is sufficient [with the (model- and
  energy-dependent) scattering time, $\tau_{\rm dis}$, assumed to have
  the appropriate value determined by $T$ and $\mu$], such that
\begin{equation}
\label{i1dis}
\bs{\cal I}_1^{\rm dis} = -\bs{j}/\tau_{\rm dis}.
\end{equation}
The generalized cyclotron frequency,
\begin{equation}
\label{ombt}
\omega_B=eBv_g^2/(c{\cal{T}}),
\end{equation}
and the vector quantity
$\bs{\cal{K}}$
\begin{equation}
\label{lj04}
\bs{\cal K}\left[\delta f\right]={\cal T}N\sum_\lambda \int\!\frac{d^2k}{(2\pi)^2}
\frac{\bs{k}}{k^2} \delta f_{\lambda\bs{k}},
\end{equation}
appear after integrating the Lorentz term. 
Substituting Eqs.~(\ref{df}) into the integral (\ref{lj04}),
one finds \cite{me1}
\begin{equation}
\label{k1}
\bs{\cal K}[\delta f]=\frac{v_gT}{2} \frac{\partial n}{\partial\mu}
\left[
\bs{h}^{(1)} \tanh\frac{x}{2} + \bs{h}^{(2)} + \bs{h}^{(3)} \frac{\cal T}{T}
\right].
\end{equation}

Since Eq.~(\ref{jeq01}) differs from that considered in
Ref.~\onlinecite{me1} by the time derivative term in the right-hand
side only, one can anticipate that the frequency-dependent dissipative
coefficients can be elucidated from the results of
Ref.~\onlinecite{me1} by adding the frequency to $\tau_{\rm dis}^{-1}$
\[
\tau_{\rm dis}^{-1} 
\quad\rightarrow\quad
\tau_{\rm dis}^{-1} \!+\! \partial/\partial t 
\quad\rightarrow\quad
\tau_{\rm dis}^{-1} \!-\! i\omega.
\]

The integrated Liouville's operator, collision integral, and Lorentz
term in Eq.~(\ref{jeq01}) were evaluated in Ref.~\onlinecite{me1}. To
the linear order in $\bs{u}$ (within linear response in
$\bs{E}$)
\begin{eqnarray}
&&
\!\!\!\!\!\!
N\!\sum_\lambda\!\!\int\!\!\frac{d^2k}{(2\pi)^2} \bs{v}_{\lambda\bs{k}} {\cal L} f^{(0)}_{\lambda\bs{k}}
\!=\!
n\frac{\partial\bs{u}}{\partial t}
+
\frac{v_g^2}{2}\bs{\nabla}n
-
\frac{v_g^2}{2}e\bs{E}\frac{\partial n}{\partial \mu}
\nonumber\\
&&
\nonumber\\
&&
\qquad\qquad\qquad\qquad\qquad\qquad
-\frac{1}{2}v_g^2\frac{e}{c}\frac{\partial n}{\partial \mu}\bs{u}\!\times\!\bs{B}.
\end{eqnarray}
\end{subequations}
Expressing the time derivative in terms of gradients with the help of
the Euler equation [i.e., Eq.~(\ref{eq1g}) without the viscous terms
  and magnetic field], one may express the integrated Liouville's
operator in terms of the gradient of the electro-chemical potential
(in the absence of temperature gradients)
\begin{widetext}
\[
N\!\sum_\lambda\!\int\!\frac{d^2k}{(2\pi)^2} \bs{v}_{\lambda\bs{k}} {\cal L} f^{(0)}_{\lambda\bs{k}}
=
v_g^2\left(\frac{2n^2}{3n_{E}}\!-\!\frac{1}{2}\frac{\partial n}{\partial \mu}\right)
\!\left(\bs{F}_{\!\bs{u}}\!-\!T\bs{\nabla}\frac{\mu}{T}\right)
-
\left(\frac{2n n_{I}}{3n_{E}}\!-\!\frac{1}{2}\frac{\partial n_{I}}{\partial \mu}
\right)T\bs{\nabla}\frac{\mu_I}{T}.
\]

Similarly to Eq.~(\ref{jeq0}) one finds the macroscopic equation for
the imbalance current. This time the kinetic equation is multiplied by
$\lambda\bs{v}_{\lambda\bs{k}}$ which upon integration yields
\begin{subequations}
\begin{equation}
\label{jIeq0}
N\!\sum_\lambda\!\lambda\!\int\!\!\frac{d^2k}{(2\pi)^2} 
\bs{v}_{\lambda\bs{k}} {\cal L} f^{(0)}_{\lambda\bs{k}}
\!=\!
\bs{\cal I}_2\left[f\right]
\!-\!
\frac{\partial\delta\bs{j}_I}{\partial t}
\!-\!
\omega_B \bs{e}_B\!\times\!\bs{\cal K}_I\left[\delta f\right]\!.
\end{equation}
In the right-hand side of Eq.~(\ref{jIeq0}), the integrated Lorentz
term in Eq.~(\ref{jIeq0}) contains the vector quantity
\begin{equation}
\label{ljI04}
\bs{\cal K}_I[\delta f]={\cal T}N\sum_\lambda\lambda \int\!\frac{d^2k}{(2\pi)^2}
\frac{\bs{k}}{k^2}\delta f_{\lambda\bs{k}}
=
\frac{v_gT}{2} \frac{\partial n}{\partial\mu}
\left[
\bs{h}^{(1)}  + \bs{h}^{(2)} \tanh\frac{x}{2} + x \bs{h}^{(3)} 
\right]\!,
\end{equation}
\end{subequations}
and ${\bs{\cal{I}}_2\left[f\right]}$ denotes the integrated collision
integral [defined similarly to Eq.~(\ref{i1def}), but with the extra
  factor of $\lambda$]. The left-hand side of Eq.~(\ref{jIeq0}) can be
evaluated similarly to that in Eq.~(\ref{jeq0}) such that \cite{me1}
\[
N\!\sum_\lambda\!\lambda\!\int\!\frac{d^2k}{(2\pi)^2} 
\bs{v}_{\lambda\bs{k}} {\cal L} f^{(0)}_{\lambda\bs{k}}
=
v_g^2\!\left(\frac{2nn_I}{3n_{E}}\!-\!\frac{1}{2}\frac{\partial n_I}{\partial \mu}\right)
\!\left(\bs{F}_{\!\bs{u}}\!-\!T\bs{\nabla}\frac{\mu}{T}\right)
-
\left(\frac{2n_{I}^2}{3n_{E}}\!-\!\frac{1}{2}\frac{\partial n_{I}}{\partial \mu}
\right)T\bs{\nabla}\frac{\mu_I}{T}.
\]


Finally, the energy current $\bs{j}_E$ is proportional to the momentum
density and hence the Liouville's operator acting on the local
equilibrium distribution function yields zero. The integrated
collision integral in the equation for the energy current vanishes due
to momentum conservation. As a result, the macroscopic equation for the
energy current contains only the Lorentz and disorder scattering terms
\begin{equation}
\label{ljE}
0 = - \frac{\delta\bs{j}_E}{\tau_{\rm dis}} - \frac{\partial\delta\bs{j}_E}{\partial t}
- \omega_B \frac{{\cal T}}{T}\bs{e}_B\!\times\!\delta\bs{j}.
\end{equation}

Combining the three macroscopic currents into a vector, one can write
all three macroscopic equations as a single equation in the matrix
form \cite{me1}
\begin{subequations}
\label{eq}
\begin{eqnarray}
\label{djeq2}
\widehat{\textswab{M}}_n
\begin{pmatrix}
  \bs{F}_{\!\bs{u}}-T\,\bs{\nabla}\displaystyle\frac{\mu}{T} \cr
  T\,\bs{\nabla}\displaystyle\frac{\mu_I}{T} \cr
  0
\end{pmatrix}
=
\left[
\frac{\alpha_g^2T^2}{2{\cal T}^2}
\widehat{\textswab{T}}
+
\frac{\pi}{\cal T}\left(\frac{1}{\tau_{\rm dis}}\!+\!\frac{\partial}{\partial t}\right)
\widehat{\textswab{M}}_h
\right]
\widehat{\textswab{M}}_h^{-1}
\!
\begin{pmatrix}
  \delta\bs{j} \cr
  \delta\bs{j}_I \cr
  \delta\bs{j}_E/T 
\end{pmatrix}
+
\pi
\frac{\omega_B}{\cal T}
\widehat{\textswab{M}}_K
\widehat{\textswab{M}}_h^{-1}
\bs{e}_B\!\times\!
\begin{pmatrix}
  \delta\bs{j} \cr
  \delta\bs{j}_I \cr
  \delta\bs{j}_E/T 
\end{pmatrix}\!,
\end{eqnarray}
where the dimensionless matrix $\widehat{\textswab{M}}_n$ describes
the integrated Liouville's operators in the equations for the
quasiparticle currents,
\begin{equation}
\label{mdim}
\begin{pmatrix}
\frac{2n^2}{3n_{E}}\!-\!\frac{1}{2}\frac{\partial n}{\partial \mu} &
-\frac{2n n_{I}}{3n_{E}}\!+\!\frac{1}{2}\frac{\partial n_{I}}{\partial \mu} & 0\cr
\frac{2n n_{I}}{3n_{E}}\!-\!\frac{1}{2}\frac{\partial n_{I}}{\partial \mu} &
-\frac{2n_{I}^2}{3n_{E}}\!+\!\frac{1}{2}\frac{\partial n}{\partial \mu} & 0\cr
0 & 0 & 0
\end{pmatrix}\!
=
- \frac{1}{2}\frac{\partial n}{\partial \mu} \, \widehat{\textswab{M}}_n,
\qquad
\widehat{\textswab{M}}_n=
\begin{pmatrix}
1\!-\!\frac{2\tilde{n}^2}{3\tilde{n}_E}\frac{T}{\cal T} &
\frac{\tilde{n}}{3\tilde{n}_E}\!\left[x^2\!+\!\frac{\pi^2}{3}\right]\!\frac{T}{\cal T}
\!-\!\frac{xT}{\cal T} & 0 \cr
\frac{xT}{\cal T} \!-\!
\frac{\tilde{n}}{3\tilde{n}_E}\!\left[x^2\!+\!\frac{\pi^2}{3}\right]\!\frac{T}{\cal T} & 
\frac{1}{6\tilde{n}_E}\!\left[x^2\!+\!\frac{\pi^2}{3}\right]^2\!\frac{T}{\cal T}\!-\!1 & 0 \cr
0 & 0 & 0 
\end{pmatrix}\!,
\end{equation}
the dimensionless matrix $\widehat{\textswab{M}}_K$ describes the
integrated Lorentz terms,
\begin{equation}
\label{mk}
\widehat{\textswab{M}}_K
=
\begin{pmatrix}
\tanh\frac{x}{2} & 1 & \frac{\cal T}{T} \cr
1 & \tanh\frac{x}{2} & x \cr
\frac{\cal T}{T} & x & 2\tilde{n}
\end{pmatrix}\!,
\end{equation}
the integrated collision integral due to electron-electron interaction
is expressed in terms of the ``scattering rates'' \cite{me1},
\begin{equation}
\label{iee}
\begin{pmatrix}
  \bs{\cal I}^{ee}_1 \cr
  \bs{\cal I}^{ee}_2 \cr
  0
\end{pmatrix}
=
-\frac{v_gT}{2}\frac{\partial n}{\partial \mu}
\begin{pmatrix}
  \tau_{11}^{-1} & \tau_{12}^{-1} & 0 \cr
  \tau_{12}^{-1} & \tau_{22}^{-1} & 0 \cr
  0 & 0 & 0
\end{pmatrix}\!
\begin{pmatrix}
\bs{h}^{(1)} \cr
\bs{h}^{(2)} \cr
\bs{h}^{(3)}
\end{pmatrix},
\qquad
\begin{pmatrix}
  \tau_{11}^{-1} & \tau_{12}^{-1} & 0 \cr
  \tau_{12}^{-1} & \tau_{22}^{-1} & 0 \cr
  0 & 0 & 0
\end{pmatrix}
=
\frac{\alpha_g^2}{8\pi}\frac{NT^2}{\cal T} \widehat{\textswab{T}},
\end{equation}
\end{subequations}
and finally $\tau_{\rm dis}^{-1}$ describing disorder scattering
enters the equation together with the time derivative (as explained
above).

Solving the linear equations (\ref{eq}) by usual methods \cite{me1}, one finds
for ${\mu_I=0}$ and neglecting temperature gradients
\begin{subequations}
\label{sigmagen}
\begin{eqnarray}
\label{sbom}
&&
\begin{pmatrix}
  \delta\bs{j} \cr
  \delta\bs{j}_I \cr
  \delta\bs{j}_E/T 
\end{pmatrix}
=
\widehat{\textswab{M}}_h
\left(
1\!+\!\widehat{\textswab{S}}_{xx}^{-1}\widehat{\textswab{S}}_{xy}
\widehat{\textswab{S}}_{xx}^{-1}\widehat{\textswab{S}}_{xy}
\right)^{-1}
\widehat{\textswab{S}}_{xx}^{-1}\widehat{\textswab{M}}_n
\begin{pmatrix}
  \bs{F}_{\!\bs{u}}-\bs{\nabla}\mu \cr
  0 \cr
  0
\end{pmatrix}
\\
&&
\nonumber\\
&&
\qquad\qquad\qquad\qquad\qquad\qquad
-
\widehat{\textswab{M}}_h
\left(
1\!+\!\widehat{\textswab{S}}_{xx}^{-1}\widehat{\textswab{S}}_{xy}
\widehat{\textswab{S}}_{xx}^{-1}\widehat{\textswab{S}}_{xy}
\right)^{-1}
\widehat{\textswab{S}}_{xx}^{-1}\widehat{\textswab{S}}_{xy}\widehat{\textswab{S}}_{xx}^{-1}
\widehat{\textswab{M}}_n
\bs{e}_B\!\times\!
\begin{pmatrix}
  \bs{F}_{\!\bs{u}}-\bs{\nabla}\mu \cr
  0 \cr
  0
\end{pmatrix},
\nonumber
\end{eqnarray}
where
\begin{equation}
\label{sxx}
\widehat{\textswab{S}}_{xx}
=
\frac{\alpha_g^2T^2}{2{\cal T}^2}
\widehat{\textswab{T}}
+
\frac{\pi}{\cal T}\left(\frac{1}{\tau_{\rm dis}}-i\omega\right)
\widehat{\textswab{M}}_h,
\qquad
\widehat{\textswab{S}}_{xy}
=
\pi\frac{\omega_B}{\cal T}
\widehat{\textswab{M}}_K.
\end{equation}
\end{subequations}
Eq.~(\ref{sigmagen}) provides a closed expression for the dissipative
corrections to the three macroscopic currents in the hydrodynamic
picture of electronic transport in graphene. This result generalizes
the earlier static solution \cite{me1} providing the dissipative
contribution to optical conductivity. While at charge neutrality the
dissipative correction represents the total electric current, see
Eq.~(\ref{djs}), away from the Dirac point one has to take into
account the time-dependent solution of the hydrodynamic equations, the
Navier-Stokes equation (\ref{eq1g}) and continuity equations. Such
solution can be obtained in a closed analytic form in the two limiting
cases, either close to charge neutrality or in the degenerate regime.
These two cases will be considered in detail in the remainder of this
paper. In the most general case (e.g., for ${\mu\sim{T}}$) the
hydrodynamic equations have to be solved numerically. Such analysis is
beyond the scope of the present paper and will be discussed elsewhere.

\clearpage

\end{widetext}

\section{Optical conductivity in zero magnetic field}

Consider first graphene in the absence of magnetic field. Then the
dissipative correction to the electric current is given by the first
line in Eq.~(\ref{sigmagen}), which now simplifies to
\begin{subequations}
\begin{equation}
\label{sg1}
\begin{pmatrix}
  \delta\bs{j} \cr
  \delta\bs{j}_I \cr
  \delta\bs{j}_E/T 
\end{pmatrix}
=
\widehat{\textswab{M}}_h
\textswab{S}_{xx}^{-1}\widehat{\textswab{M}}_n
\begin{pmatrix}
  e\bs{E}-\bs{\nabla}\mu \cr
  0 \cr
  0
\end{pmatrix}\!.
\end{equation}
The corresponding contribution to conductivity can then be
formally expressed as
\begin{equation}
\label{sigma0}
\delta\sigma(\omega)
=
\begin{pmatrix}
  1 & 0 & 0
\end{pmatrix}
\widehat{\textswab{M}}_h
\textswab{S}_{xx}^{-1}\widehat{\textswab{M}}_n
\begin{pmatrix}
  1 \cr
  0 \cr
  0
\end{pmatrix}\!.
\end{equation}
\end{subequations}
In addition to Eq.~(\ref{sigma0}), the full electrical conductivity in
graphene comprises also the ``hydrodynamic'' contribution, see the
first term in the electric current, Eq.~(\ref{djdef}). In order to
distinguish between the two, I will refer to Eq.~(\ref{sigma0}) as the
``kinetic'' contribution \cite{fog}.

\subsection{Hydrodynamic contribution to optical conductivity}

The hydrodynamic contribution to the optical conductivity should be
obtained by solving the Navier-Stokes equation (\ref{eq1g}). Focusing
on the homogeneous (${q=0}$) solution, I may write Eq.~(\ref{eq1g}) in
the form [${\delta\bs{j}_E(\bs{B}\!=\!0)=0}$]
\begin{subequations}
\label{eqs0q}  
\begin{equation}
W\partial_t\bs{u}
+
\bs{u} \partial_t P 
=
v_g^2 
en\bs{E}
-
\frac{W\bs{u}}{\tau_{{\rm dis}}}.
\end{equation}
Looking for homogeneous solutions of the continuity equations
(\ref{cen1}) and (\ref{cene1}), I conclude that
\begin{equation}
\partial_tn=0, \qquad \partial_tn_E=0,
\end{equation}
hence the time derivative of pressure in the Navier-Stokes equation
vanishes as well (${P=n_E/2}$). Then the equation takes the form
\begin{equation}
\label{nseq0q}
W\bs{u}\left(\tau_{\rm dis}^{-1}-i\omega\right) = v_g^2 en\bs{E},
\end{equation}
with the solution
\begin{equation}
\label{nuh}
n\bs{u} = \frac{v_g^2 n^2}{W} \frac{e\bs{E}}{\tau_{\rm dis}^{-1}-i\omega}.
\end{equation}
\end{subequations}
The resulting hydrodynamic contribution to the optical conductivity is
given by \cite{har}
\begin{equation}
\label{sigmah}
\sigma_h(\omega) = \frac{e^2v_g^2 n^2}{W} \frac{1}{\tau_{\rm dis}^{-1}-i\omega}.
\end{equation}

\subsubsection{Hydrodynamic contribution to conductivity in the degenerate regime}

In the degenerate regime, ${x\gg1}$, the enthalpy and carrier density
can be expressed in terms of the chemical potential,
\begin{subequations}
\label{flqs}
\begin{equation}
\label{wfl}
W=3P=\frac{3n_E}{2}=\frac{\mu^3}{\pi v_g^2}\left(1\!+\!\frac{\pi^2}{x^2}\right),
\end{equation}
\begin{equation}
\label{nfl}
n = \frac{\mu^2}{\pi v_g^2}\left(1\!+\!\frac{\pi^2}{3x^2}\right).
\end{equation}
\end{subequations}
Substituting these expressions into Eq.~(\ref{sigmah}) and keeping the
leading correction only, one finds
\begin{equation}
\label{shfl}
\sigma_h(\omega) = \frac{e^2\mu}{\pi} \frac{1}{\tau_{\rm dis}^{-1}-i\omega}
\left(1-\frac{\pi^2}{3x^2}\right).
\end{equation}
In the limit ${x\rightarrow\infty}$ the result (\ref{shfl}) represents
the total optical conductivity, since the dissipative correction
(\ref{sg1}) vanishes, see below.

\subsubsection{Hydrodynamic contribution to conductivity near charge neutrality}

At charge neutrality, ${n=0}$, the electrical current (\ref{djs}) is
completely determined by the dissipative correction (\ref{sg1}). For
nonzero, but small carrier density, ${x\ll1}$, the carrier and energy
densities can be expanded as
\begin{subequations}
\label{dpqs}  
\begin{equation}
\label{ndp}
n = \frac{NT^2}{2\pi v_g^2} 2x\ln2 + {\cal O}(x^3),
\end{equation}
\begin{equation}
\label{nedp}
n_E = \frac{NT^3}{2\pi v_g^2} \left[ 3\zeta(3) + 2x^2\ln2\right] + {\cal O}(x^4).
\end{equation}
\end{subequations}
This yields the following solution (to the leading order in ${x\ll1}$)
of the Navier-Stokes equation (\ref{nseq0q})
\[
n\bs{u} = \frac{16\ln^22}{9\pi\zeta(3)} x^2 T
\frac{e\bs{E}}{\tau_{\rm dis}^{-1}\!-\!i\omega} + {\cal O}(x^4).
\]
The resulting contribution to the optical conductivity is
\begin{equation}
\label{shdp}
\sigma_h(\omega) = \frac{e^2T}{\pi} \frac{16\ln^22}{9\zeta(3)} \frac{x^2}{\tau_{\rm dis}^{-1}-i\omega}
+ {\cal O}(x^4).
\end{equation}
The numerical coefficient in Eq.~(\ref{shdp}) is of order unity,
\[
\frac{16\ln^22}{9\zeta(3)} \approx 0.71.
\]

Close to charge neutrality, the result (\ref{shdp}) is subleading to
the kinetic contribution to conductivity, which I discuss next.

\subsection{Kinetic contribution to optical conductivity}

Consider now the second contribution to conductivity,
Eq.~(\ref{sigma0}), which stems from the dissipative correction to the
electric current (\ref{sg1}). Since the general expression is not
transparent enough, I now consider the limiting cases.

\subsubsection{Optical conductivity at charge neutrality}

At charge neutrality, ${x=0}$, the matrices $\widehat{\textswab{M}}_h$
and $\widehat{\textswab{M}}_n$ simplify,
\begin{subequations}
\label{dpms}
\begin{equation}
\label{m0}
\widehat{\textswab{M}}_h
=
\begin{pmatrix}
1 & 0 & 0 \cr
0 & 1 & \frac{\pi^2}{6\ln2} \cr
0 & \frac{\pi^2}{6\ln2} & \frac{9\zeta(3)}{2\ln2}
\end{pmatrix}\!,
\quad
\widehat{\textswab{M}}_n
=
\begin{pmatrix}
1 & 0 & 0 \cr
0 & -\delta & 0 \cr
0 & 0 & 0
\end{pmatrix}\!,
\end{equation}
where $\zeta(z)$ is the Riemann's zeta function and 
\[
\delta = 1-\frac{\pi^4}{162\zeta(3)\ln2}
\approx 0.28.
\]
The matrix of the scattering rates simplifies as well since
${\tau_{12}^{-1}(\mu\!=\!0)=0}$ \cite{hydro1,me1}. As a result, 
the matrix $\textswab{S}_{xx}$ has the block-diagonal form
\begin{equation}
\label{sxx0}
\widehat{\textswab{S}}_{xx}
\!=\!
\frac{\pi}{2T\ln2}\!
\left[\!
\begin{pmatrix}
  \tau_{11}^{-1} & 0 & 0 \cr
  0 & \tau_{22}^{-1} & 0 \cr
  0 & 0 & 0
\end{pmatrix}
\!+\!
\left(\frac{1}{\tau_{\rm dis}}\!-\!i\omega\right)
\widehat{\textswab{M}}_h\!
\right]\!\!,
\end{equation}
\end{subequations}
where the electric current decouples from the other two. In the
absence of temperature gradients the correction $\delta\bs{j}_I$
vanishes [also ${\delta\bs{j}_E(\bs{B}\!=\!0)=0}$]. This can be seen
from Eq.~(\ref{sg1}) where the source term (i.e., the external
electric field) is only present in the electric current sector. The
conclusion about the energy current can be reached already from
Eq.~(\ref{ljE}).

At zero frequency, the resulting electrical conductivity is given by
Eq.~(\ref{sq}):
\[
e\delta\bs{j} 
= 
\frac{2\ln2}{\pi} e^2T\left(\frac{1}{\tau_{11}}\!+\!\frac{1}{\tau_{\rm dis}}\right)^{\!\!-1}\!\bs{E}
\underset{\tau_{\rm dis}\rightarrow\infty}{\longrightarrow}
\sigma_Q\bs{E}.
\]
The coefficient ${\cal A}$ in Eq.~(\ref{sq}) can be found by
evaluating the scattering rate numerically
\cite{kash,schutt,mfss,hydro1,me1}
\begin{equation}
\label{t11a}
{\cal A} = 2(\ln2) \alpha_g^2 T\tau_{11} \approx 0.12.
\end{equation}
The explicit expression for $\tau_{11}$ is given in
Appendix~\ref{apptau}. Note, that the above value of ${\cal A}$ was
calculated with the unscreened Coulomb potential. At charge
neutrality, this is a reasonable approximation for weak coupling
(i.e., ${\alpha_g\ll1}$). For larger values of $\alpha_g$, which may
be more experimentally relevant, screening leads to a quantitatively
significant change in ${\cal A}$.

\begin{figure}[t]
\centerline{\includegraphics[width=0.9\columnwidth]{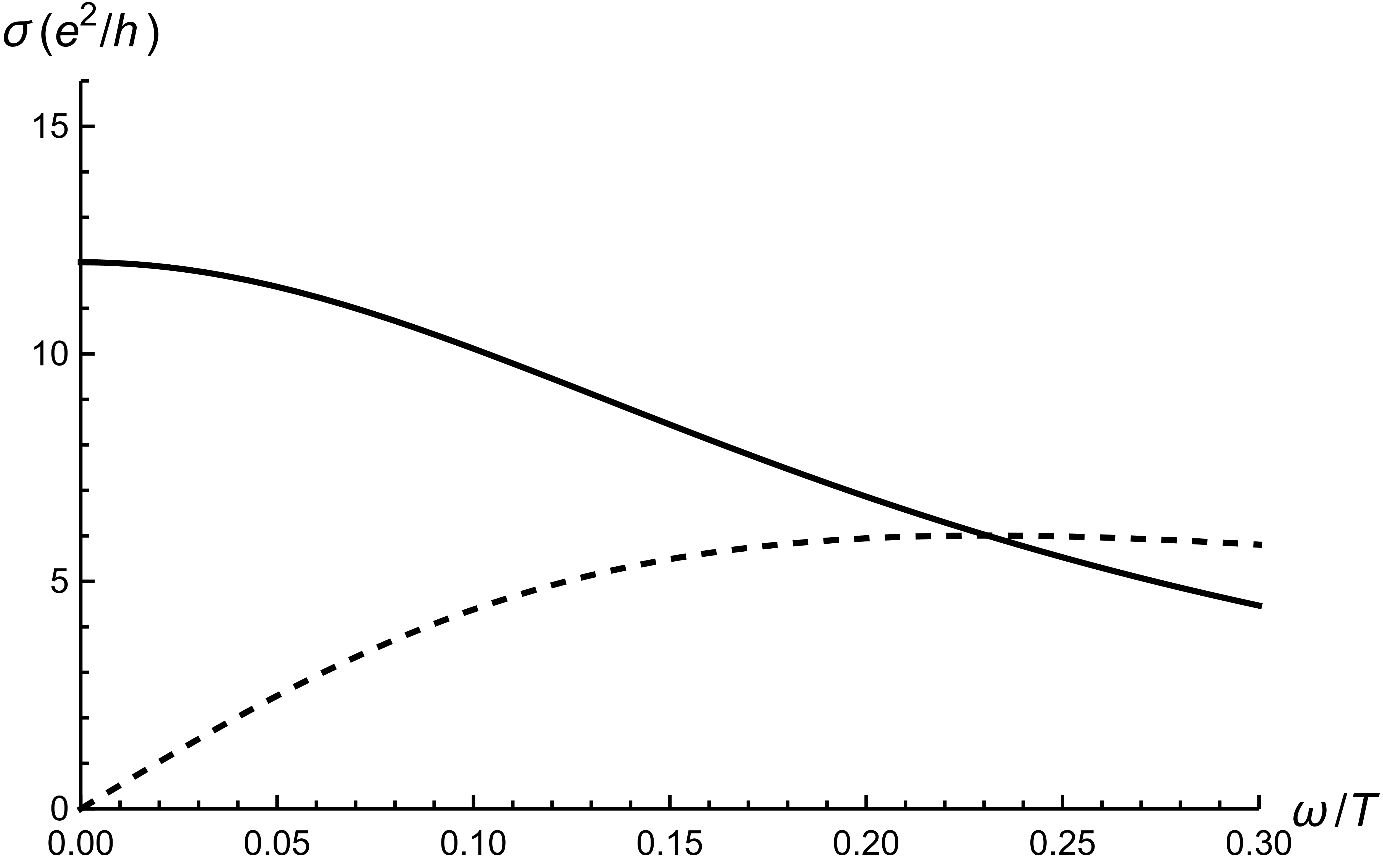}
}
\caption{Optical conductivity in graphene at charge neutrality. The
  real and imaginary parts of Eq.~(\ref{som0}) are shown by the solid
  and dashed curves, respectively. The shown dependence appears to
  agree with the experimental data of Ref.~\onlinecite{gal} (see
  Fig.~3D of that reference). The curves were calculated with
  ${\alpha_g=0.23}$ and ${\tau_{\rm dis}^{-1}=0.8\,}$THz, the values
  taken from Ref.~\onlinecite{gal}.}
\label{fig:som0}
\end{figure}

Keeping the nonzero frequency in Eq.~(\ref{sxx0}), I find the optical
conductivity in graphene at charge neutrality \cite{fog,gal},
\begin{equation}
\label{som0}
\sigma(\omega; \mu\!=\!0)=\frac{2\ln2}{\pi}
\frac{e^2T}{-i\omega\!+\!\tau_{\rm dis}^{-1}\!+\!\tau_{11}^{-1}},
\end{equation}
where $\tau_{11}$ is the same as in Eq.~(\ref{t11a}). This result is
illustrated in Fig.~\ref{fig:som0}.

\subsubsection{Optical conductivity in the degenerate regime}

Deep in the degenerate (or ``Fermi-liquid'') regime, i.e. for
${\mu\gg{T}}$, the matrices in Eq.~(\ref{sigma0}) become
degenerate. In particular, the matrix $\widehat{\textswab{M}}_h$
determining the relation (\ref{djsm}) between the dissipative
corrections to macroscopic currents and the nonequilibrium
distribution function takes the form (up to exponentially small
corrections)
\begin{equation}
\label{mhdfl}
\widehat{\textswab{M}}_h=
\begin{pmatrix}
1 & 1 & x\left[1\!+\!\frac{\pi^2}{3x^2}\right] \cr
1 & 1 & x\left[1\!+\!\frac{\pi^2}{3x^2}\right] \cr
x\left[1\!+\!\frac{\pi^2}{3x^2}\right] & x\left[1\!+\!\frac{\pi^2}{3x^2}\right] &
x^2\left[1\!+\!\frac{\pi^2}{x^2}\right]
\end{pmatrix}\!,
\end{equation}
where I have used the asymptotic expressions
\[
\tilde{n}\approx\tilde{n}_I=\frac{x^2}{2}\!+\!\frac{\pi^2}{6},
\qquad
\tilde{n}_E\approx\frac{x^3}{6}\!+\!\frac{\pi^2x}{6},
\qquad
\frac{\cal T}{T}\approx x.
\]
The first of these equalities follows from the fact that in the
degenerate regime only one band of carriers contributes to transport
(the contribution of the other is exponentially suppressed; this is
the reason why the two first rows in $\widehat{\textswab{M}}_h$ are
identical). In the limit $x\rightarrow\infty$ the factor
$\pi^2/x^2\rightarrow0$ can be neglected. Then the third row of the
matrix $\widehat{\textswab{M}}_h$ is proportional to the first two,
such that all three dissipative corrections in Eq.~(\ref{djsm}) are
proportional to each other. Now, regardless of the value of $x$,
${\delta\bs{j}_E(B\!=\!0)=0}$, see Eq.~(\ref{ljE}), and hence we
conclude
\begin{subequations}
\begin{equation}
\label{djsfllim}
\delta\bs{j}(x\rightarrow\infty)=\delta\bs{j}_I(x\rightarrow\infty)
\approx\delta\bs{j}_E=0.
\end{equation}
Note, that this result cannot be obtained from Eq.~(\ref{sigma0}) with
the asymptotic form of $\widehat{\textswab{M}}_h$ and other matrices
since in this limit the matrix $\textswab{S}_{xx}$ is degenerate (with
or without the power law corrections). To find corrections to
Eq.~(\ref{djsfllim}) one has to either to keep track of exponentially
small corrections to Eq.~(\ref{mhdfl}) or, alternatively, disregard
the imbalance mode (since ${\delta\bs{j}=\delta\bs{j}_I}$ with
exponential accuracy). Inverting the remaining ${2\times2}$ matrix, one
finds
\begin{equation}
\label{seefl}
\delta\sigma(\omega) =
\frac{\pi^3e^2\mu}{3x^2\left[3x^2\tau_{11}^{-1}+\pi^2(\tau_{\rm dis}^{-1}-i\omega)\right]}.
\end{equation}
The scattering rate $\tau_{11}^{-1}$ in the degenerate regime has the
asymptotic form (neglecting screening; see
Appendix~\ref{apptaufl})
\begin{equation}
\label{t11fl}
\tau_{11}^{-1}(x\gg1) \approx \frac{4\pi}{3} \alpha_g^2 \frac{T^2}{\mu} =
\frac{4\pi}{3} \alpha_g^2 \frac{\mu}{x^2}.
\end{equation}
\end{subequations}

Combining both contributions, Eqs.~(\ref{seefl}) and (\ref{shfl}), I find
the optical conductivity in graphene as
\begin{equation}
\label{somfl}
\sigma(\omega) = \frac{e^2\mu}{\pi} \frac{1}{\tau_{\rm dis}^{-1}\!-\!i\omega}
\!\left[1\!-\!
\frac{\tau_{11}^{-1}}
     {(3x^2/\pi^2)\tau_{11}^{-1}\!+\!\tau_{\rm dis}^{-1}\!-\!i\omega}
     \right]\!,
\end{equation}
where the leading contribution is determined by disorder scattering
only. The kinetic contribution to conductivity, $\delta\sigma$, is
suppressed as $x^{-2}$. Moreover, the frequency dependence in
$\delta\sigma$ becomes important only when it is comparable to
${(3x^2/\pi^2)\tau_{11}^{-1}\gg\tau_{11}^{-1}}$, see below.

\subsubsection{Optical conductivity close to the Dirac point}

Close to charge neutrality, i.e. for nonzero, but small carrier
densities, ${x\ll1}$, one finds corrections to
Eq.~(\ref{som0}). Expanding the matrices $\widehat{\textswab{M}}_{h(n)}$
around the Dirac point, one finds
\[
\widehat{\textswab{M}}_{h(n)} = \widehat{\textswab{M}}_{h(n)}(x\!=\!0) 
+ \delta\widehat{\textswab{M}}_{h(n)} + {\cal O}(x^3), 
\]
where ${\widehat{\textswab{M}}_{h(n)}(0)}$ are given in
Eq.~(\ref{m0}), while the leading-order corrections are given by
\[
\delta\widehat{\textswab{M}}_h
=x
\begin{pmatrix}
0 & \frac{1}{2\ln2} & 2 \cr
\frac{1}{2\ln2} & 0 & \frac{x}{2\ln2}\left[1-\frac{\pi^2}{24\ln2}\right] \cr 
2 & \frac{x}{2\ln2}\left[1\!-\!\frac{\pi^2}{24\ln2}\right] &
3x\left[1\!-\!\frac{3\zeta(3)}{16\ln^22}\right]
\end{pmatrix}\!,
\]
and
\begin{eqnarray*}
&&
\!\!\!\!\!\delta\widehat{\textswab{M}}_n
=
x
\\
&&
\\
&&
\times
\!\!
\begin{pmatrix}
- \frac{8\ln2}{27\zeta(3)}x & \frac{4\pi^2\!\ln2-27\zeta(3)}{2\ln2} & 0 \cr
-\frac{4\pi^2\!\ln2-27\zeta(3)}{2\ln2} & 
\frac{\pi^2x}{27\zeta(3)\ln2}\!
\left[1\!-\!\frac{\pi^2}{48\ln2}\!-\!\frac{\pi^2\!\ln2}{9\zeta(3)}\right] & 0 \cr 
0 & 0 & 0
\end{pmatrix}\!\!.
\end{eqnarray*}
The matrix $\textswab{S}_{xx}$ can be expanded in the same way, using
the expansion of the scattering rates given in
Appendix~\ref{apptaudp}. To the leading order, one finds
\[
\textswab{S}_{xx} = \textswab{S}_{xx}(x\!=\!0) + \delta \textswab{S}_{xx} + {\cal O}(x^3),
\]
where $\textswab{S}_{xx}(x\!=\!0)$ is given by Eq.~(\ref{sxx0}) and
\[
\delta \textswab{S}_{xx}
=
\frac{\alpha_g^2}{8\ln^22}\delta\widehat{\textswab{T}}
+
\frac{\pi}{2T\ln2}\left(\tau_{\rm dis}^{-1}\!-\!i\omega\right)\delta\widehat{\textswab{M}}_h,
\]
with
\[
\delta\widehat{\textswab{T}}
=
x\begin{pmatrix}
\frac{x}{t_{11}^{(2)}}\!-\!\frac{1}{8\ln2}\frac{x}{t_{11}^{(0)}} 
& 1/t_{12}^{(1)} & 0 \cr
1/t_{12}^{(1)} 
& \frac{x}{t_{22}^{(2)}}\!-\!\frac{1}{8\ln2}\frac{x}{t_{22}^{(0)}} & 0 \cr
0 & 0 & 0
\end{pmatrix},
\]
see Eqs.~(\ref{tausdp}) for notations.

\begin{figure}[t]
\centerline{\includegraphics[width=0.9\columnwidth]{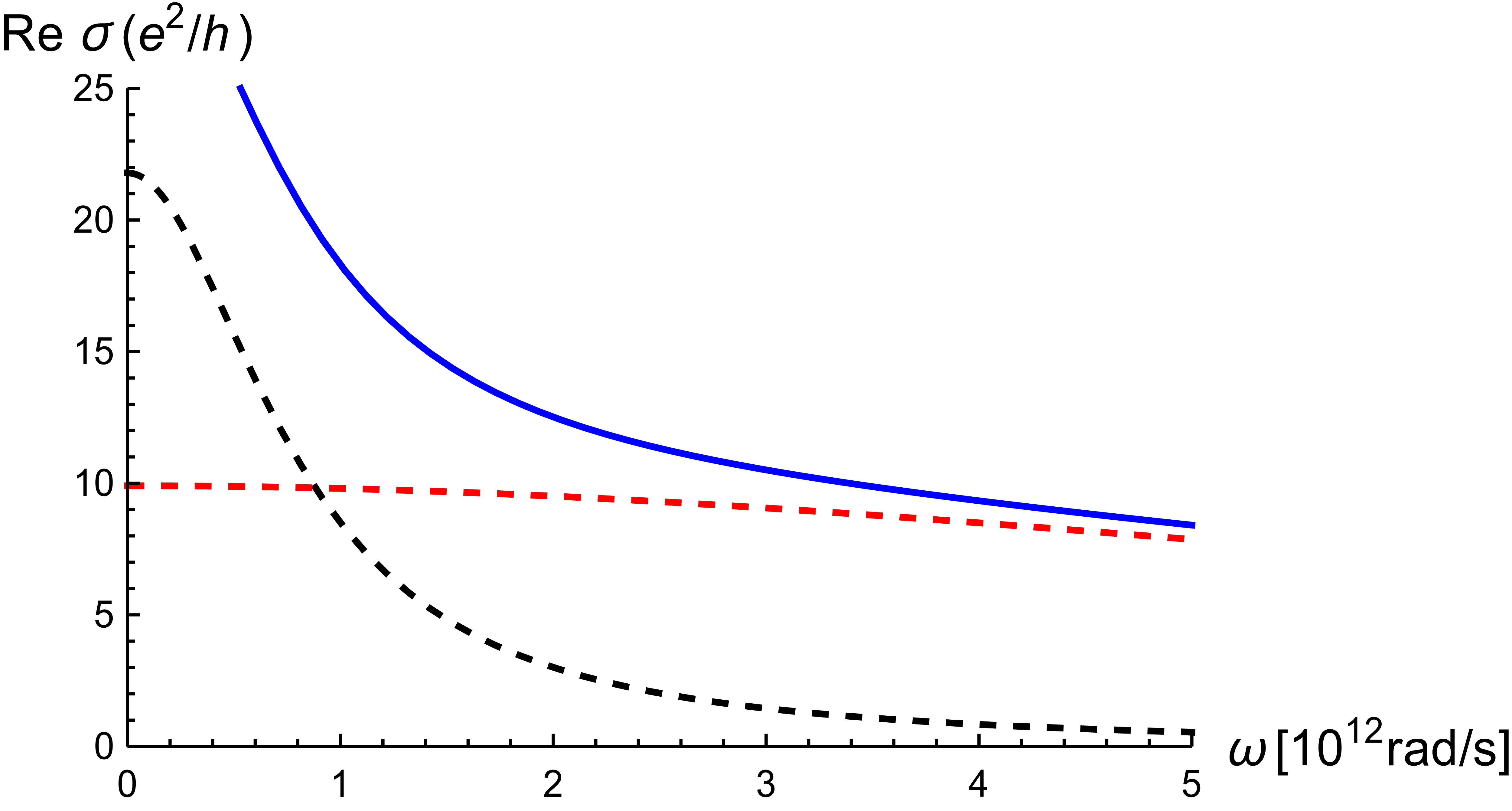}
}
\caption{Optical conductivity in weakly doped graphene at
  ${n=0.08}\,$cm$^{-12}$ (or ${E_F=33}\,$meV, the value used in
  Ref.~\onlinecite{gal}, cf. Fig.~4B of that reference). The almost
  flat red dashed curve shows the real part of the kinetic
  contribution (\ref{sigma0}), while the black dashed curve shows the
  real part of the hydrodynamic contribution (\ref{shdp}). The real
  part of the full electrical conductivity (i.e., the sum
  ${\delta\sigma+\sigma_h}$) is shown by the solid blue curve. The
  curves were calculated with ${\alpha_g=0.23}$, ${T=298}\,$K, and
  ${\tau_{\rm dis}^{-1}=0.8\,}$THz, the values taken from
  Ref.~\onlinecite{gal}.}
\label{fig:somhk}
\end{figure}

Substituting the above expansions into Eq.~(\ref{sigma0}), one finds
(see Appendix~\ref{appsomdp} for details)
\begin{subequations}
\label{somdpexp}
\begin{equation}  
\delta\sigma(\omega) = \sigma(\omega; \mu\!=\!0) + x^2 \delta\sigma^{(2)}(\omega)
+ {\cal O}(x^3),
\end{equation}
\begin{eqnarray}
\label{ds2}
&&
\delta\sigma^{(2)} =
\frac{\gamma_{11}e^2T}{-i\omega\!+\!\tau_{\rm dis}^{-1}\!+\!\tau_{11}^{-1}(0)}
+
\frac{\gamma_{12}e^2T}{-i\omega\!+\!\tau_{\rm dis}^{-1}\!+\!\gamma_{13}\tau_{22}^{-1}(0)}
\nonumber\\
&&
\nonumber\\
&&
\qquad
+
\frac{e^2T\left[
\gamma_{31} \tau_{22}^{-1}(0)
+
\tilde\gamma_{32} \left(-i\omega\!+\!\tau_{\rm dis}^{-1}\right)
-
\tilde\gamma_{33}/\tau_{12}^{(1)}\right]}
{\left[-i\omega\!+\!\tau_{\rm dis}^{-1}\!+\!\tau_{11}^{-1}(0)\right]
 \left[-i\omega\!+\!\tau_{\rm dis}^{-1}\!+\!\gamma_{13}\tau_{22}^{-1}(0)\right]}
\nonumber\\
&&
\nonumber\\
&&
\qquad
+
\frac{e^2T}{2\pi^2}
\frac{1/\tau_{11}^{(2)}-1/[8\tau_{11}(0)\ln2]}
     {\left[-i\omega\!+\!\tau_{\rm dis}^{-1}\!+\!\tau_{11}^{-1}(0)\right]^2},
\end{eqnarray}
where
\begin{equation}
\gamma_{11}\approx0.075, \quad \gamma_{12}\approx0.66, \quad \gamma_{13}\approx3.59,
\end{equation}
\begin{equation}
\gamma_{31}\approx0.81,
\end{equation}
\begin{equation}
\tilde\gamma_{32}=\gamma_{32}+\gamma_{41}\approx0.91,
\quad
\tilde\gamma_{33}=\gamma_{42}-\gamma_{33}\approx0.102.
\end{equation}
\end{subequations}

The total optical conductivity in the vicinity of the Dirac point is
given by the sum of the leading term (\ref{som0}), the hydrodynamic
contribution (\ref{shdp}), and the above correction (\ref{ds2}). The
two terms are compared in Fig.~\ref{fig:somhk}.

\subsection{Comparison with the existing literature}

Optical conductivity in graphene was already studied in several
publications, so it is worthwhile to compare the above calculations to
the previously known results.

The authors of the pioneering paper, Ref.~\onlinecite{har}, used
general hydrodynamic arguments to find the optical conductivity at a
generic charge density in the form
\[
\sigma_{xx} = \sigma_Q + \sigma_h(\omega),
\]
where $\sigma_Q$ is the dissipative constant (\ref{sq}) and the
hydrodynamic contribution is given (up to the choice of normalization)
by Eq.~(\ref{sigmah}). The specific value of $\sigma_Q$ for graphene
was calculated in Refs.~\onlinecite{kash,mfss,mfs,schutt}.

Conductivity at the Dirac point in the absence of disorder was studied
in detail in Ref.~\onlinecite{mfss}. The result of this paper is the
same as Eq.~(\ref{som0}) without the disorder scattering rate. The
electron-electron scattering rate estimated in Ref.~\onlinecite{mfss}
\[
\tau_{11}^{-1}\rightarrow\kappa\alpha_g^2T, 
\quad
\kappa = 3.646,
\]
perfectly agrees with Eq.~(\ref{tau11dp}), where the numerical
prefactor is ${\kappa=3.8}$. Note that this value was obtained for
unscreened Coulomb interaction.

Optical conductivity beyond charge neutrality was reported in
Ref.~\onlinecite{mfs}. Here the authors combined the result of
Ref.~\onlinecite{mfss} with the hydrodynamic contribution previously
reported in Ref.~\onlinecite{har}. In addition, the effect of disorder
scattering was studied not only in the limit of weak disorder, but
also for strong disorder, where the hydrodynamic approach is no longer
valid. For weak disorder, the authors of Ref.~\onlinecite{mfs}
reported an expansion in the disorder strength.

The effect of disorder was also studied in Ref.~\onlinecite{schutt} at
charge neutrality. The reported zero-frequency conductivity agrees
with Eq.~(\ref{som0}) for weak disorder, ${T\tau_{\rm dis}\gg1}$. For
stronger disorder, the authors of Ref.~\onlinecite{schutt} took into
account the effect of disorder scattering on the excitation spectrum
in graphene which is beyond the scope of the present paper.

More recently, the interpolation formula (\ref{s0}) was suggested in
Ref.~\onlinecite{fog} (see also Ref.~\onlinecite{svin}) with the
coefficient $A_1$ coinciding with that in Eq.~(\ref{sigmah}) and the
coefficient $A_2$ given by
\[
A_2\rightarrow e^2\left[\frac{\cal T}{\pi} - \frac{v_g^2n^2}{W}\right],
\]
interpolating between the two limits:
Eq.~(\ref{som0}) at charge neutrality and Eq.~(\ref{shfl}) in the
degenerate regime. In the former case, ${n=0}$ (with ${A_1=0}$)
and ${{\cal T}=2T\ln2}$, while in the latter case,
${{\cal{T}}=\pi{v}_g^2n^2/W=\mu}$, with ${A_2=0}$.

The interpolation formula Eq.~(\ref{s0}) was used to interpret the
result of recent measurements of the optical conductivity in graphene
reported in Ref.~\onlinecite{gal}. At charge neutrality, the
experimental data appears to confirm the result Eq.~(\ref{som0}) with
${\tau_{\rm dis}\gg\tau_{11}}$, see Fig.~\ref{fig:som0}, providing a
reasonable (in line with previous measurements \cite{sav,bas})
estimate for $\alpha_g$ in real graphene. Data away from charge
neutrality were measured at rather low charge densities, where the
scattering rates Eq.~(\ref{tauij_z_dl}) exhibit only small deviations
from their values at ${\mu=0}$ (in agreement with the results of
Ref.~\onlinecite{gal}).  Now, the kinetic contribution to optical
conductivity Eq.~(\ref{sigma0}) has a Lorentzian-like shape as a
function of $\omega$ [see Eqs.~(\ref{sigmagen}), (\ref{som0}),
  (\ref{seefl}), and (\ref{somdpexp})]. Hence, Eq.~(\ref{sigma0}) can
be approximated by the second term in Eq.~(\ref{s0}) by choosing the
appropriate (phenomenological) values for $\tau_{ee}$ and $\tau_{\rm
  dis}$ (as was done in Ref.~\onlinecite{gal}, where the data was
fitted by Eq.~(\ref{s0}) with the electronic temperature and
$\tau_{\rm dis}$ used as fitting parameters). Note, that this
procedure neglects renormalization of $v_g$ due to electron-electron
interaction \cite{shsch,luc,mfss}.

\begin{figure}[t]
\centerline{\includegraphics[width=0.9\columnwidth]{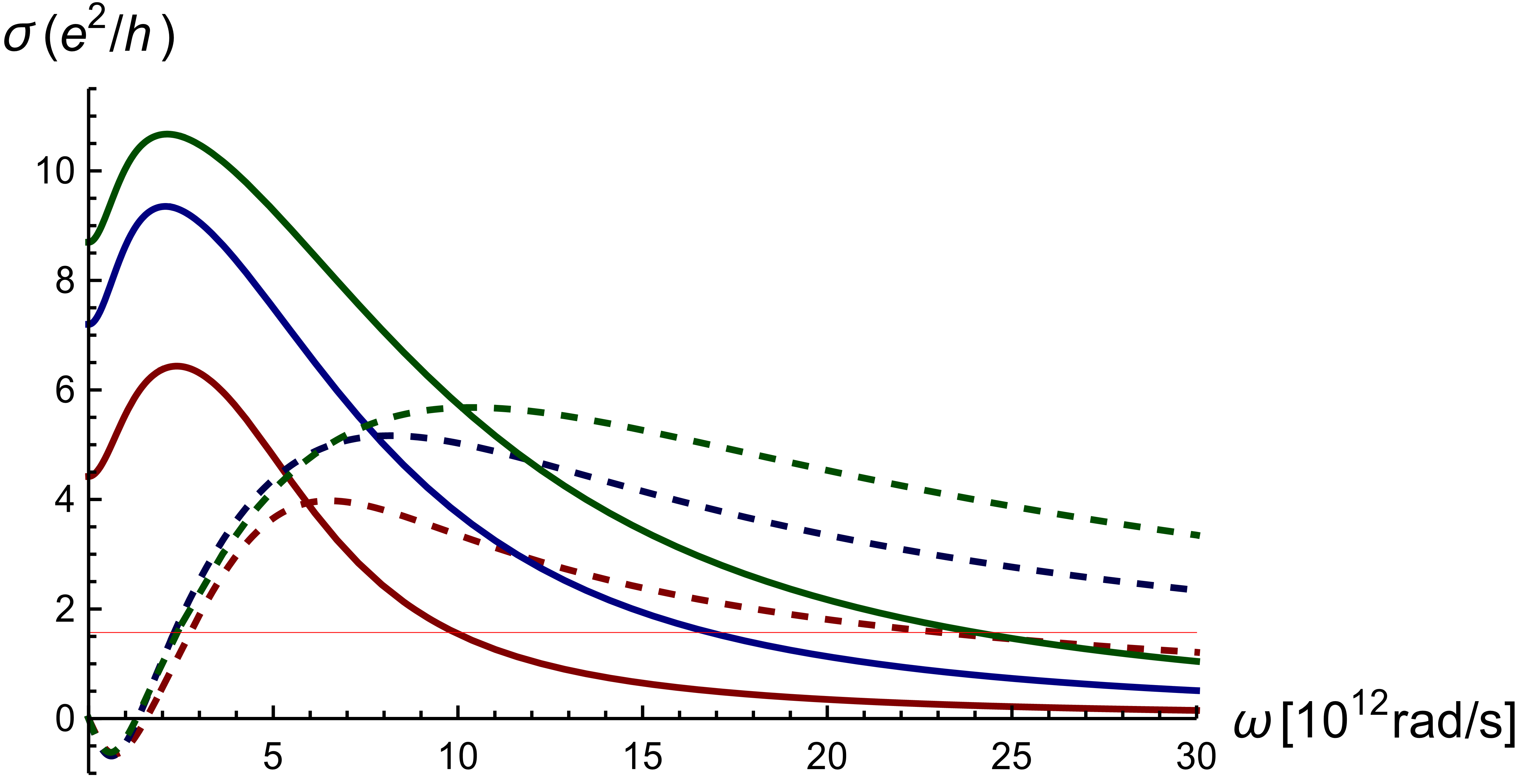}
}
\caption{Optical conductivity in graphene at charge neutrality in weak
  magnetic field, ${B\!=\!0.1}\,$T. The real and imaginary parts of
  Eq.~(\ref{sigmaBom0}) are shown by the solid and dashed curves, with
  red, blue, and green (in ascending order for solid curves)
  corresponding to ${T=100,200,300\,}$K, respectively. The curves were
  calculated using the parameter values described in the main text
  based on the data of Ref.~\onlinecite{gal}. The horizontal red line
  shows the universal conductivity Eq.~(\ref{snc}) of free electrons
  in graphene in the high-frequency collisionless regime
  \cite{mish,teber,ssh,julia2,geimop,geimop1,julia1} (actually
  observed at much higher frequencies).}
\label{fig:som0B}
\end{figure}

\begin{figure}[t]
\centerline{\includegraphics[width=0.88\columnwidth]{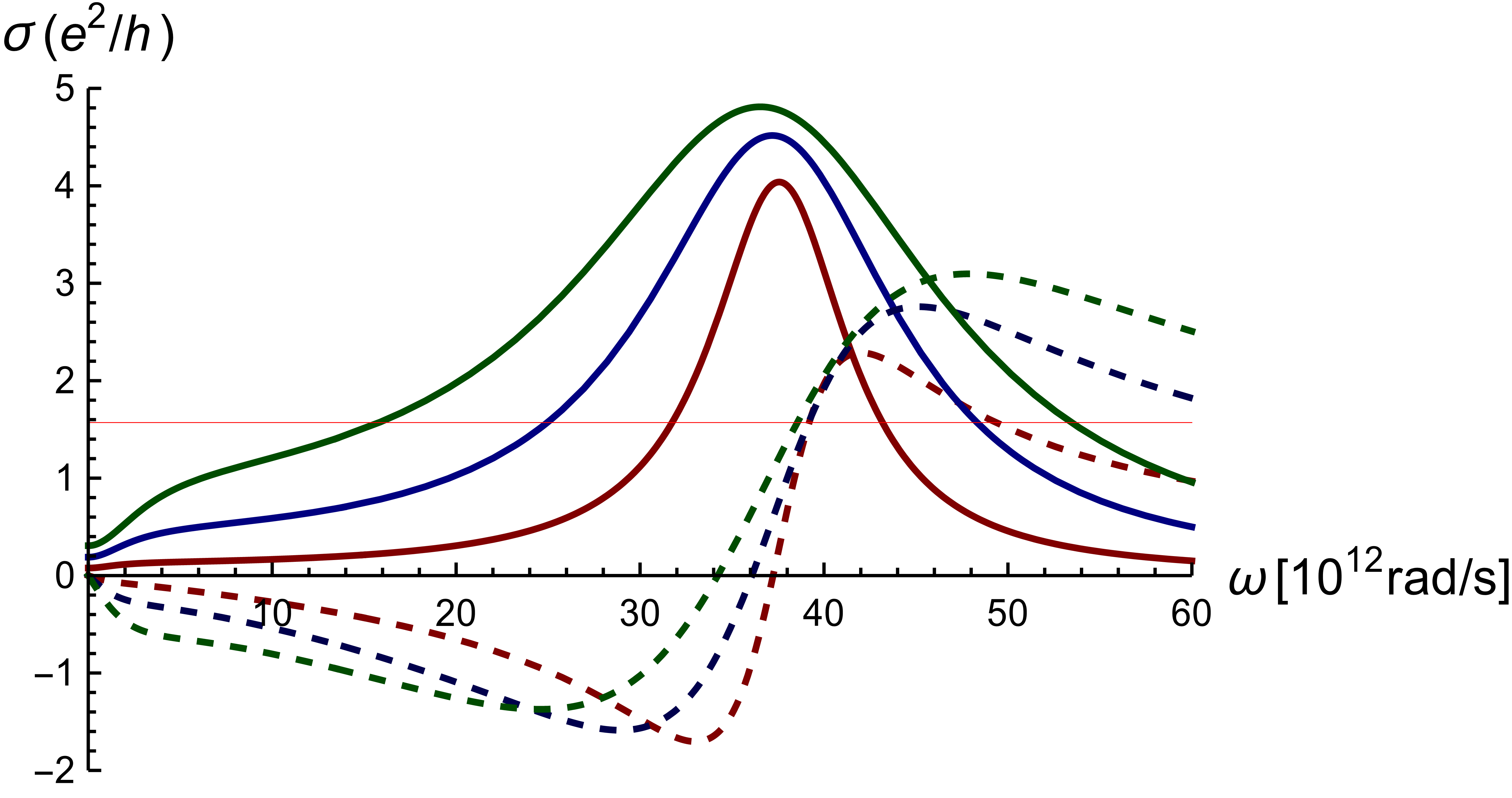}
}
\caption{Optical conductivity in graphene at charge neutrality at
  ${B\!=\!1}\,$T. The real and imaginary parts of
  Eq.~(\ref{sigmaBom0}) are shown by the solid and dashed curves, with
  red, blue, and green (in ascending order for solid curves)
  corresponding to ${T=100,200,300\,}$K, respectively. The curves were
  calculated using the same parameter values as in
  Fig.\ref{fig:som0B}. The horizontal red line shows the universal
  conductivity Eq.~(\ref{snc}) of free electrons in graphene in the
  high-frequency collisionless regime
  \cite{mish,teber,ssh,julia2,geimop,geimop1,julia1} (actually
  observed at much higher frequencies).}
\label{fig:som0B1}
\end{figure}

\section{Magnetoconductivity in graphene}

Consider now the effect of classical magnetic fields on the optical
conductivity in graphene. The classical approach is justified at high
enough temperatures where quantized transport is smeared out.

The general expression for the dissipative corrections to macroscopic
currents is given by Eqs.~(\ref{sigmagen}). In the presence of
the magnetic field, these corrections become ``entangled'' with the
hydrodynamic velocity due to the Lorentz contribution to
$\bs{F}_{\bs{u}}$.

\subsection{Magnetoconductivity at charge neutrality}

The dissipative correction to the energy current was defined through
Eq.~(\ref{ljE}). Substituting this relation into the Navier-Stokes
equation (\ref{eq1g}) at charge neutrality, one finds a stationary and
uniform solution, ${\bs{u}=0}$. Evaluating all matrices in
Eqs.~(\ref{sigmagen}), one has to consider not only Eqs.~(\ref{dpms}),
but also the matrix (\ref{mk}). At ${x\!=\!0}$ I find
\begin{equation}
\label{mkdp}
\widehat{\textswab{M}}_K
=
\begin{pmatrix}
0 & 1 & 2\ln2 \cr
1 & 0 & 0 \cr
2\ln2 & 0 & 0
\end{pmatrix}\!.
\end{equation}
Instead of multiplying the matrices in the general solution
(\ref{sbom}), it appears to be more instructive to write
Eq.~(\ref{djeq2}) explicitly at ${x\!=\!0}$ and proceed with solving
the resulting equations.

Using Eq.~(\ref{m0}), the left-hand side of Eq.~(\ref{djeq2}) takes
the form
\[
\widehat{\textswab{M}}_n
\begin{pmatrix}
  e\bs{E} \cr
  0 \cr
  0
\end{pmatrix}
=
\begin{pmatrix}
  e\bs{E} \cr
  0 \cr
  0
\end{pmatrix}.
\]
The longitudinal part of the right-hand side of Eq.~(\ref{djeq2}) is
given by multiplying Eq.~(\ref{sxx0}) by
$\widehat{\textswab{M}}_h^{-1}$, see Eq.~(\ref{m0}),
\[
\frac{\pi}{2T\ln2}\!
\left[\!
\begin{pmatrix}
  \tau_{11}^{-1} & 0 & 0 \cr
  0 & \tau_{22}^{-1}\delta^{-1} & -\tau_{22}^{-1} \frac{\pi^2}{27\zeta(3)\delta} \cr
  0 & 0 & 0
\end{pmatrix}
\!+\!
\frac{1}{\tau_{\rm dis}}\!-\!i\omega
\right]\!,
\]
while the Hall part is given by
\[
\frac{\pi\omega_B}{2T\ln2}\!
\begin{pmatrix}
0 & \delta_1 & -\delta_2 \cr
1 & 0 & 0 \cr
2\ln2 & 0 & 0
\end{pmatrix}
\!\bs{e}_B\!\times\!\!
\begin{pmatrix}
  \delta\bs{j} \cr
  \delta\bs{j}_I \cr
  \delta\bs{j}_E/T 
\end{pmatrix}\!,
\]
where
\[
\delta_1 = \frac{1}{\delta}\!-\!\frac{2\pi^2\ln2}{27\zeta(3)\delta}
\approx2.08,
\quad
\delta_2 = \frac{\pi^2\!-\!12\ln^22}{27\zeta(3)\delta}
\approx0.45.
\]
As a result, the three equations for macroscopic currents simplify
admitting a simple analytical solution. This solution is analogous to
the solution of the stationary equations leading to the zero-frequency
conductivity, see Refs.~\onlinecite{hydro0,me1}.

The third such equation stems from Eq.~(\ref{ljE}) and allows one to
express the correction to the energy current, $\delta\bs{j}_E$, in
terms of the correction to the electric current, $\delta\bs{j}$,
\begin{subequations}
\label{eqsdp}
\begin{equation}
\label{jEeqdp}
\delta\bs{j}_E
=
-2(\ln2)\frac{\omega_BT}{\tau_{\rm dis}^{-1}\!-\!i\omega}\bs{e}_B\!\times\!\delta\bs{j}.
\end{equation}
The second equation determines the correction to the imbalance current
\begin{equation}
\label{jIeqdp}
\delta\bs{j}_I
=
-\frac{\omega_B}{\tau_{\rm dis}^{-1\!}-\!i\omega}\!
\left[1\!-\!\frac{\delta_1\tau_{22}^{-1}}{\delta^{-1}\tau_{22}^{-1}\!+\!\tau_{\rm dis}^{-1}\!-\!i\omega}\right]
\bs{e}_B\!\times\!\delta\bs{j}.
\end{equation}
The first equation determines the electrical current
\begin{eqnarray}
\label{jeqdp}
&&
e\bs{E} = \frac{\pi}{2T\ln2} \left(\tau_{11}^{-1}\!+\!\tau_{\rm dis}^{-1}\!-\!i\omega\right)
\delta\bs{j}
\\
&&
\nonumber\\
&&
\qquad\qquad
+
\frac{\pi\omega_B}{2T\ln2} \left(\delta_1 \bs{e}_B\!\times\!\delta\bs{j}_I
-
\delta_2\bs{e}_B\!\times\!\delta\bs{j}_E/T\right).
\nonumber
\end{eqnarray}
Both of the corrections (\ref{jEeqdp}) and (\ref{jIeqdp}) are
orthogonal to the electric current, hence
${\delta\bs{j}\|\bs{E}}$. This means that no orthogonal component of
the current is generated, in other words, there is no classical Hall
effect in graphene at charge neutrality, as expected. Substituting the
relations (\ref{jEeqdp}) and (\ref{jIeqdp}) into Eq.~(\ref{jeqdp}), I
find
\end{subequations}
\begin{widetext}
\begin{equation}
\label{sigmaBom0}
\sigma_{xx}(\omega) = \frac{2\ln2}{\pi}
\frac{e^2T\left(\tau_{\rm dis}^{-1}-i\omega\right)\!
            \left(\delta^{-1}\tau_{22}^{-1}\!+\!\tau_{\rm dis}^{-1}-i\omega\right)}
{\left(\tau_{\rm dis}^{-1}-i\omega\right)\!
  \left(\delta^{-1}\tau_{22}^{-1}\!+\!\tau_{\rm dis}^{-1}-i\omega\right)\!
  \left(\tau_{11}^{-1}\!+\!\tau_{\rm dis}^{-1}\!-\!i\omega\right)
\!+\!\omega_B^2\left[\delta_3\tau_{22}^{-1}\!+\!\delta_4\left(\tau_{\rm dis}^{-1}-i\omega\right)
\right]},
\end{equation}
\end{widetext}
where
\[
\delta_3 \!=\! \frac{\delta_1}{\delta}\!-\!\delta_1^2\!-\!2\frac{\delta_2}{\delta}\ln2
\approx0.88,
\quad
\delta_4 \!=\! \delta_1\!-\!2\delta_2\ln2 \approx 1.45.
\]
At ${\bs{B}=0}$, the expression (\ref{sigmaBom0}) reproduces
Eq.~(\ref{som0}) as it should. At ${\omega=0}$, I recover the
positive, parabolic magnetoresistance previously found in
Refs.~\onlinecite{mfs,hydro0,me1}.

Let me now estimate the relative value of various parameters in
Eq.~(\ref{sigmaBom0}) using the data of
Ref.~\onlinecite{gal}. Measurements at charge neutrality yield the
following value of the coupling constant, ${\alpha_g\!=\!0.23}$. Using
Eqs.~(\ref{tausdp}), this lead to the following estimates of
the scattering rates at a typical temperature $267\,$K
\[
\tau_{11}^{-1}\approx 7.35 \, {\rm THz},
\qquad
\tau_{22}^{-1}\approx 4.17 \, {\rm THz}.
\]
The disorder scattering rate at $267\,$K was estimated as
\[
\tau_{\rm dis}^{-1} \approx 0.8  \, {\rm THz}.
\]
Finally, evaluating Eq.~(\ref{ombt}) at a reference magnetic field,
${B=1}\,$T, the frequency $\omega_B$ can be estimated as
\[
\omega_B=\frac{|e|v_g^2B}{2cT\ln2}\approx 27.9 \, {\rm THz}
\frac{B}{1 \, {\rm T}} \frac{300 \, {\rm K}}{T},
\]
such that at $267\,$K and $0.1\,$T one finds
${\omega_B\approx3.13}\,$THz.  For comparison, $267\,$K
${\rightarrow34.96}\,$THz, so the hydrodynamic picture is fully
justified.

At higher frequencies and magnetic fields the hydrodynamic picture
breaks down \cite{mfs} in the sense that the local equilibrium
(underlying the traditional derivation of the hydrodynamic equations
\cite{dau10,hydro1,me1,luc}) cannot be formed. At the same time, the
kinetic (Boltzmann) equation has a much wider range of
applicability. Within linear response, the kinetic equation can be
integrated to obtain the macroscopic equations for quasiparticle
currents \cite{hydro0,pol17} without referring to local
equilibrium. Remarkably, these linear response equations coincide with
the macroscopic current equations (\ref{jeq0}), (\ref{jIeq0}), and
(\ref{ljE}), see Ref.~\onlinecite{me1} for more details. On one hand,
this means that the results for linear response quantities, e.g. the
electrical conductivity, obtained within the hydrodynamic theory
coincide with the results of the linear response theory. On the other
hand, the result (\ref{sigmaBom0}) which has not been worked out
within the linear response theory has a wider regime of applicability
than the concept of local equilibrium. Extending the range of
frequencies and magnetic fields in Eq.~(\ref{sigmaBom0}) one arrives
at the resonance-like picture illustrated in Fig.~\ref{fig:som0B1}.
Note that at these frequencies one typically assumes the system to be
in the collisionless regime
\cite{mish,teber,ssh,julia2,geimop,geimop1,julia1}, where the optical
conductivity is given by the universal value (up to weak interaction
corrections)
\begin{subequations}
\label{snc}
\begin{equation}
\sigma_U(\omega) = \frac{\pi e^2}{2 h}
\left[1+{\cal C}_\sigma \alpha_g(\omega)\right], 
\end{equation}
where the renormalized coupling constant $\alpha_g(\omega)$ and the
numerical coefficient are given by
\begin{equation}
\label{ar}
\alpha_g(\omega) = \frac{\alpha_g}{1\!+\!\frac{\alpha_g}{4}\ln\frac{\cal D}{\omega}},
\quad
{\cal C}_\sigma \approx 0.01.
\end{equation}
\end{subequations}
Physically, the universal value (\ref{snc}) is due to interband
transitions and is hence beyond the semiclassical Boltzmann equation
discussed in this paper. One typically assumes that the solutions to
the Boltzmann equation yield the conductivity decaying with frequency,
such that in the high-frequency regime Eq.~(\ref{snc}) dominates. In
the hydrodynamic regime this is illustrated in Fig.~\ref{fig:som0B},
where the universal value (\ref{snc}) is shown by the horizontal red
line. Extending the kinetic theory beyond the hydrodynamic regime
leads to the cyclotron resonance at high frequencies where the
solution to the Boltzmann equation dominates over Eq.~(\ref{snc}), see
Fig.~\ref{fig:som0B1}.

\subsection{Magnetoconductivity in the degenerate regime}

In the degenerate or Fermi-liquid regime only one band contributes to
electronic transport and hence ${\bs{j}=\bs{j}_I}$ with exponential
accuracy (see above). At the same time, away from charge neutrality
the dissipative corrections (\ref{sigmagen}) depend on the
hydrodynamic velocity $\bs{u}$ and hence the ``hydrodynamic'' and
``kinetic'' contributions to conductivity become entangled. In this
case, it seems more transparent to consider the macroscopic equations
for the electrical and energy currents explicitly. As shown in
Ref.~\onlinecite{me1}, these equations are exactly the same as the
macroscopic linear response equations considered in
Ref.~\onlinecite{hydro0}. Similar equations have been derived in
Ref.~\onlinecite{pol17}.

The equation for the energy current is just the Navier-Stokes equation
(\ref{eq1g}). For a homogeneous flow the gradient terms and the time
derivative of pressure vanish, see the above derivation of
Eq.~(\ref{nseq0q}), such that the equation takes the form
\begin{subequations}
\label{fleqs}
\begin{equation}
\label{nseqB0q}
\bs{j}_E\left(\tau_{\rm dis}^{-1}-i\omega\right) = v_g^2 en\bs{E}
+ v_g^2\frac{e}{c}\bs{j}\!\times\!\bs{B}.
\end{equation}
The equation for the electric current is given by
Eq.~(\ref{jeq0}). For a homogeneous flow this equation takes the form
\begin{equation}
\label{jeqB0q}
\bs{j}\left(\tau_{\rm dis}^{-1}-i\omega\right) = \frac{1}{2}v_g^2\frac{\partial n}{\partial\mu}e\bs{E}
+
\omega_B\bs{\cal K}\!\times\!\bs{e}_{B}
+
\bs{\cal I}_1^{ee}.
\end{equation}
\end{subequations}

To the leading order \cite{hydro0}, ${\bs{j}_E\approx\mu\bs{j}}$,
${\bs{\cal{K}}\approx\bs{j}}$, and ${\bs{\cal I}_1^{ee}\rightarrow0}$.
In this case, the two equations (\ref{fleqs}) are identical yielding
the standard Drude-like results \cite{mard}
\begin{subequations}
\label{fl0res}
\begin{equation}
\label{sflres}
\sigma_{xx} = \frac{e^2\mu\tau_{\rm dis}}{\pi} 
\frac{1-i\omega\tau_{\rm dis}}{(1-i\omega\tau_{\rm dis})^2+\omega_B^2\tau_{\rm dis}^2},
\end{equation}
\begin{equation}
\label{shflres}
\sigma_{xy} = -\frac{e^2\mu\tau_{\rm dis}}{\pi} 
\frac{\omega_B\tau_{\rm dis}}{(1-i\omega\tau_{\rm dis})^2+\omega_B^2\tau_{\rm dis}^2}.
\end{equation}
\end{subequations}
Here the electric field is taken to be directed along the $x$-axis,
the magnetic field -- along the $z$-axis. Note that the electric field
is assumed to be oscillating, ${E\sim\exp(-i\omega{t})}$, and the
magnetic field is assumed to be static. For ${\omega=0}$, I recover
the standard magnetoconductivity \cite{mard,hydro0}.

At ${\bs{B}=0}$ the results (\ref{fl0res}) yield the Drude
conductivity given by Eq.~(\ref{shfl}) where one has to set
${x\rightarrow\infty}$. The leading correction to the Drude result is
given by Eq.~(\ref{somfl}) where both the correction in
Eq.~(\ref{shfl}) and the kinetic contribution (\ref{seefl}) are taken
into account.

The origin of these corrections is in the differences between the two
equations (\ref{fleqs}). Indeed, setting ${\bs{B}=0}$ and subtracting
Eq.~(\ref{nseqB0q}) from Eq.~(\ref{jeqB0q}), one finds
\begin{eqnarray*}
&&
\left(\bs{j}-\frac{\bs{j}_E}{\mu}\right)\left(\tau_{\rm dis}^{-1}-i\omega\right)
=
v_g^2\left(\frac{1}{2}\frac{\partial n}{\partial\mu}-n\right)e\bs{E}+
\bs{\cal I}_1^{ee}
\\
&&
\\
&&
\qquad\qquad\qquad\qquad\qquad
=
- \frac{\pi^2}{3x^2} \frac{\mu}{\pi} e\bs{E}+
\bs{\cal I}_1^{ee},
\end{eqnarray*}
where I used the equilibrium carrier density and compressibility in
the limit ${x\rightarrow\infty}$, see Appendix~\ref{leqs}, keeping
only the leading power law term.

In the same limit, the collision integral $\bs{\cal I}_1^{ee}$ can be
expressed in terms of the two currents as
\begin{eqnarray*}
&&
\bs{\cal I}_1^{ee} = -\frac{1}{2} v_gT\frac{\partial n}{\partial\mu}
\tau_{11}^{-1} \bs{h}^{(1)}
\\
&&
\\
&&
\qquad\quad
=
\frac{3x^2}{\pi^2} \frac{\tau_{11}^{-1}}{1\!-\!\frac{\pi^2}{3x^2}}
\left[
\left(1\!+\!\frac{\pi^2}{x^2}\right)\bs{j}
-
\left(1\!+\!\frac{\pi^2}{3x^2}\right)\frac{\bs{j}_E}{\mu}
\right]\!.
\end{eqnarray*}
Note that although Eq.~(\ref{djsm}) relates the function
$\bs{h}^{(1)}$ to the dissipative corrections $\delta\bs{j}$ and
$\delta\bs{j}_E$, the hydrodynamic part of the macroscopic currents
does not contribute to the collision integral as can be seen from the
above expression by direct substitution.

Finally, the difference in the Lorentz terms in Eqs.~(\ref{fleqs}) can
be found by evaluating the quantity $\bs{\cal K}$ to the subleading
order
\[
\bs{\cal K} = \frac{1}{1\!-\!\frac{\pi^2}{3x^2}}
\left(2\bs{j}-\frac{\bs{j}_E}{\mu}\right).
\]

Using the above expressions, solving Eqs.~(\ref{fleqs}) is a
straightforward although tedious task. The result
can be expressed as
\begin{subequations}
\label{fl1res}
\begin{equation}
\label{sfl1res}
\sigma_{xx} = \frac{e^2\mu\tau_{\rm dis}}{\pi} 
\frac{(1-i\omega\tau_{\rm dis})\left[S_0\!+\!\frac{\omega_B^2\tau_{\rm dis}^2S_1S_2}
{(1-i\omega\tau_{\rm dis})^2}\right]}{(1-i\omega\tau_{\rm dis})^2+\omega_B^2\tau_{\rm dis}^2S_1^2},
\end{equation}
\begin{equation}
\label{shfl1res}
\sigma_{xy} = -\frac{e^2\mu\tau_{\rm dis}}{\pi} 
\frac{\omega_B\tau_{\rm dis}(S_0S_1-S_2)}{(1-i\omega\tau_{\rm dis})^2+\omega_B^2\tau_{\rm dis}^2S_1^2},
\end{equation}
\end{subequations}
where
\[
S_0=1-
\frac{\tau_{11}^{-1}[(3x^2/\pi^2)\tau_{11}^{-1}\!+\!\tau_{\rm dis}^{-1}\!-\!i\omega]}
     {[(3x^2/\pi^2)\tau_{11}^{-1}\!+\!\tau_{\rm dis}^{-1}\!-\!i\omega]^2+\omega_B^2},
\]
\[
S_1=1-
\frac{3\tau_{11}^{-1}[(3x^2/\pi^2)\tau_{11}^{-1}\!+\!\tau_{\rm dis}^{-1}\!-\!i\omega]}
     {[(3x^2/\pi^2)\tau_{11}^{-1}\!+\!\tau_{\rm dis}^{-1}\!-\!i\omega]^2+\omega_B^2},
\]
\[
S_2 = \frac{\pi^2}{3x^2}
 \frac{(\tau_{\rm dis}^{-1}\!-\!i\omega)^2}
     {[(3x^2/\pi^2)\tau_{11}^{-1}\!+\!\tau_{\rm dis}^{-1}\!-\!i\omega]^2+\omega_B^2}.
\]
In the limit ${x\rightarrow\infty}$, one finds ${S_0=S_1=1}$, ${S_2=0}$
and one recovers Eqs.~(\ref{fleqs}). For ${\bs{B}=0}$, the result (\ref{sfl1res})
coincides with Eq.~(\ref{somfl}).

\section{Discussion}

In this paper I have considered optical conductivity in graphene in
the hydrodynamic regime. Within the hydrodynamic approach, the
electric current is split into the hydrodynamic and kinetic
contributions, see Eq.~(\ref{djdef}), with the latter representing the
dissipative correction. Unlike the traditional hydrodynamics
\cite{dau6} in systems with exact momentum conservation, the
hydrodynamic equations in graphene include weak impurity
scattering. As a result, both contributions to the electric current
decay yielding Eqs.~(\ref{sigmah}) and (\ref{sigma0}).

Exactly at the neutrality point, the hydrodynamic contribution to
electric current [and hence the conductivity (\ref{sigmah})]
vanishes. In this case, the optical conductivity in graphene is given
by Eq.~(\ref{shdp}) illustrated in Fig.~\ref{fig:som0}. Assuming the
disorder scattering time in Eq.~(\ref{shdp}) to be negligibly small,
one recovers the result of Ref.~\onlinecite{mfss}. This result was
recently used in Ref.~\onlinecite{gal} to establish the experimental
value of the electron-electron scattering time and ultimately the
coupling constant, $\alpha_g$. In the experiment, one treats
$\alpha_g$ as a fitting parameter using the expression
${\tau_{11}^{-1}=C\alpha_g^2T}$, with the numerical coefficient $C$
calculated in Refs.~\onlinecite{kash,schutt,mfss,hydro1,me1} assuming
the unscreened Coulomb interaction. Such approach can be justified
\cite{mfss} by $1/N$ expansion suggesting that all interaction effects
including screening and renormalization are confined to establishing
the value of the effective coupling constant but do not significantly
alter the momentum dependence of the matrix elements of the Coulomb
interaction \cite{schutt} (in the typical momentum range contributing
to the collision integral).

At high enough doping levels, the electronic system in graphene is
degenerate and most results resemble their counterparts in the usual
Fermi liquid. In particular, the optical conductivity is given by
Eq.~(\ref{shfl}). This expression is independent of the
electron-electron scattering rate, which could be interpreted as
``restoration'' of Galilean invariance in the Fermi liquid
regime. This result is fully ``hydrodynamic'' whereas the kinetic
contribution to conductivity vanishes in this limit.

At intermediate carrier densities, the electrical conductivity is
determined by the interplay of the above two mechanisms. While the
hydrodynamic contribution keeps the simple form (\ref{sigmah}), the
general expression for the kinetic contribution, Eq.~(\ref{sigmagen}),
is extremely cumbersome. It is therefore tempting to use the
interpolation formula, Eq.~(\ref{s0}), instead. As evidenced by the
leading corrections in both limits, Eqs.~(\ref{somfl}) and
(\ref{somdpexp}), the ``exact'' result (\ref{sigma0}) differs from
Eq.~(\ref{s0}). However, the resulting optical conductivity is still
roughly Lorentzian and hence could be fit by Eq.~(\ref{s0}) using the
two scattering rates as fitting parameters. Such approach was adopted
in Ref.~\onlinecite{gal} (even the electronic temperature was
obtained from the fit).

External magnetic field further complicates the theory by entangling
the two contributions to the electric current. While not affecting the
leading-order behavior in the degenerate regime, Eq.~(\ref{fl0res}),
the leading correction to the Drude behavior, Eq.~(\ref{fl1res}), is
dominated by the coupling between the two modes. At charge neutrality,
the homogeneous electric current is still unaffected by the coupling
to the hydrodynamic modes, but this is expected to change in finite
size samples where the flow becomes inhomogeneous. Previously, this
physics was used to explain giant magnetodrag in graphene \cite{meg}
and to predict linear magnetoresistance in classically strong magnetic
fields \cite{hydro0,mr1,mrexp}.

Finally, linear response transport in electronic systems can be
described within the kinetic theory in wider parameter range as
compared to the applicability region of hydrodynamics. The latter
assumes that strong electron-electron interaction establishes local
equilibrium \cite{hydro1,me1} describing the ideal hydrodynamic
flow. In contrast, the linear response theory \cite{hydro0} assumes
that in the absence of external fields the system is in the global
equilibrium state described by the Fermi-Dirac distribution function
and characterized by vanishing currents. Here the currents appear only
as a response to the external electric field. Remarkably, linearizing
the hydrodynamic equations in graphene (in order to find the linear
response coefficients, such as electrical conductivity) yields exactly
the same macroscopic equations [e.g., Eqs.~(\ref{fleqs})] as in the
linear response theory rendering these equations more general than the
concept of local equilibrium. Consequently, the results for optical
conductivity obtained in this paper are valid in a wider frequency
range than expected from the purely hydrodynamic perspective. This
conclusion is most pronounced for the cyclotron resonance in
moderately strong magnetic fields, see Fig.~\ref{fig:som0B1}, where
the kinetic contribution to the magnetoconductivity at relatively
large frequencies dominates the universal free electron conductivity
(\ref{snc}) describing the high-frequency collisionless regime that is
not included in the semiclassical Boltzmann approach \cite{mish,ssh}.
A discussion of the quantum kinetic equation \cite{mish} including the
interband transitions responsible for Eq.~(\ref{snc}) is beyond the
scope of this paper and will be reported elsewhere.

\acknowledgments

I thank A.D. Mirlin, J. Schmalian, M. Sch\"utt, and especially
I.V. Gornyi for numerous fruitful discussions. This work was supported
by the German Research Foundation DFG within the FLAG-ERA Joint
Transnational Call (Project GRANSPORT), by the European Commission
under the EU Horizon 2020 MSCA-RISE-2019 program (Project 873028
HYDROTRONICS), and the MEPhI Academic Excellence Project, Contract
No. 02.a03.21.0005.


\appendix


\section{Local equilibrium quantities}
\label{leqs}

Under the assumption of local equilibrium, the macroscopic quantities
appearing in the hydrodynamic theory can be computed explicitly
\cite{me1}. The quasiparticle densities and the energy density have
the following form
\begin{subequations}
\label{n0s}
\begin{equation}
\label{n0}
n = n_{+}\! -\! n_{-} = \frac{T^2}{v_g^2}
\frac{g_2\left(\mu_+/T\right)\!-\!g_2\left(-\mu_-/T\right)}
     {\left(1\!-\!u^2/v_g^2\right)^{3/2}},
\end{equation}
\begin{equation}
\label{nI0_0}
n_{I} \!=\! n_{+,0} \!+\! n_{-} \!=\! \frac{T^2}{v_g^2}
\frac{g_2\left(\mu_+/T\right)\!+\!g_2\left(-\mu_-/T\right)}
     {\left(1\!-\!u^2/v_g^2\right)^{3/2}},
\end{equation}
\begin{equation}
\label{ne0}
n_{E} = 2\frac{T^3}{v_g^2}
\frac{1\!+\!u^2/(2v_g^2)}
     {\left(1\!-\!u^2/v_g^2\right)^{5/2}}
\left[g_3\left(\frac{\mu_+}{T}\right)\!+\!g_3\left(-\frac{\mu_-}{T}\right)\right]\!,
\end{equation}
\end{subequations}
where
\begin{equation}
  \label{g2}
  g_2\left(\frac{\mu}{T}\right) = - \frac{N}{2\pi} {\rm Li}_2\left(-e^{\mu/T}\right),
\end{equation}
\begin{equation}
\label{g3}
g_3\left(\frac{\mu}{T}\right) = - \frac{N}{2\pi} {\rm Li}_3\left(-e^{\mu/T}\right),
\end{equation}
with ${\rm Li}_n$ being the polylogarithm. It is convenient to express
the densities (\ref{n0s}) in the dimensionless form
\begin{equation}
n = \frac{NT^2}{2\pi v_g^2} \tilde{n}, 
\quad
n_I = \frac{NT^2}{2\pi v_g^2} \tilde{n}_I,
\quad
n_E = \frac{NT^3}{\pi v_g^2} \tilde{n}_E.
\end{equation}
In the simplest case considered in this paper, ${\mu_\pm=\mu}$, the
dimensionless imbalance density simplifies to
\begin{equation}
\label{nI0}
\tilde{n}_I = \frac{x^2}{2}+\frac{\pi^2}{6}, 
\qquad
x = \frac{\mu}{T}.
\end{equation}
Similarly, the compressibilities (for ${\mu_\pm=\mu}$) are given by
\begin{equation}
\frac{\partial n}{\partial\mu} =\frac{N{\cal T}}{2\pi v_g^2},
\quad
\frac{\partial n_I}{\partial\mu} =\frac{N\mu}{2\pi v_g^2},
\quad
{\cal T} = 2T\ln\left[2\cosh\frac{x}{2}\right].
\end{equation}
The two thermodynamic quantities in the hydrodynamic theory, the pressure
and enthalpy are given by
\begin{subequations}
\label{hqsg}
\begin{equation}
\label{phyd}
P = n_{E} \frac{1-u^2/v_g^2}{2+u^2/v_g^2},
\end{equation}
\begin{equation}
\label{whyd}
W=n_{E}+P = \frac{3 n_{E}}{2+u^2/v_g^2}.
\end{equation}
\end{subequations}
They can be used to determine the stress-energy tensor
\begin{equation}
\label{pihyd}
\Pi_{E}^{\alpha\beta} \!=\!
P\delta_{\alpha\beta} + v_g^{-2}Wu_\alpha u_\beta,
\end{equation}
and the energy current (proportional to the momentum density $\bs{n}_{\bs{k}}$)
\begin{equation}
\label{jehyd}
\bs{j}_{E} = v_g^2 \bs{n}_{\bs{k}} = W\bs{u}.
\end{equation}
This relation is the key feature of the hydrodynamic description of
the electronic system in graphene showing that it is the energy and
not electric current that is described by the hydrodynamic flow. The
quasiparticle currents are determined by the corresponding densities
\begin{equation}
\bs{j} = n \bs{u},
\qquad
\bs{j}_{I} = n_{I} \bs{u}.
\end{equation}
Unlike the energy current, the quasiparticle currents are not
conserved in electron-electron collisions and hence acquire the
dissipative corrections (\ref{djdef}). Furthermore, the energy current
can be relaxed by disorder scattering and hence acquires a dissipative
correction of its own.

\clearpage

\begin{widetext}

\section{Collision integral}
\label{apptau}

The integrated collision integrals (\ref{iee}) are expressed in terms
of the following scattering rates \cite{me1}
\begin{subequations}
  \label{taus_c}
  \begin{equation}
    \label{tau_vv}
    \frac{1}{\tau_{11}} = \pi^2 \alpha_g^2NT \left[\frac{NT}{v_g^2\partial n_0/\partial\mu}\right]
\int\frac{d^2Q}{(2\pi)^2}\frac{dW}{2\pi}
\frac{|\widetilde{U}|^2}{\sinh^2W}
    \left(
    Y_{00}Y_{11}\!-\! Y_{01}^2  
    \right),
  \end{equation}
  \begin{equation}
    \label{tau_vs}
    \frac{1}{\tau_{12}} = \pi^2 \alpha_g^2NT \left[\frac{NT}{v_g^2\partial n_0/\partial\mu}\right]
\int\frac{d^2Q}{(2\pi)^2}\frac{dW}{2\pi}
\frac{|\widetilde{U}|^2}{\sinh^2W}
    \left(
      Y_{00}Y_{12}\!-\! Y_{02}Y_{01}
    \right),
  \end{equation}
  \begin{equation}
    \label{tau_ss}
    \frac{1}{\tau_{22}} = \pi^2 \alpha_g^2NT \left[\frac{NT}{v_g^2\partial n_0/\partial\mu}\right]
\int\frac{d^2Q}{(2\pi)^2}\frac{dW}{2\pi}
\frac{|\widetilde{U}|^2}{\sinh^2W}
    \left(
      Y_{00}Y_{22}\!-\! Y_{02}^2
    \right),
  \end{equation}
\end{subequations}
with
\begin{subequations}
  \label{ys_c}
  \begin{equation}
  \label{y00i}
  Y_{00}(\omega,\bs{q}) = \frac{1}{4\pi}
  \left[
    \frac{\theta(|\Omega|\leqslant1)}{\sqrt{1\!-\!\Omega^2}}\,
    {\cal Z}_0^>[I_1]
    +
    \frac{\theta(|\Omega|\geqslant1)}{\sqrt{\Omega^2\!-\!1}}\,
    {\cal Z}_0^<[I_1]
    \right]\!,
\end{equation}
\begin{equation}
  \label{y01i}
  Y_{01}(\omega,\bs{q}) = -\frac{1}{2\pi}
  \left[
    \theta(|\Omega|\leqslant1)\sqrt{1\!-\!\Omega^2}\,
    {\cal Z}_2^>[I]
    +
    \theta(|\Omega|\geqslant1)\sqrt{\Omega^2\!-\!1}\,
    {\cal Z}_2^<[I]
    \right]\!,
\end{equation}
\begin{equation}
  \label{y02i}
  Y_{02}(\omega,\bs{q}) = \frac{1}{2\pi}
  \left[
    \theta(|\Omega|\leqslant1)\sqrt{1\!-\!\Omega^2}\,
    {\cal Z}_2^>[I_1]
    +
    \theta(|\Omega|\geqslant1)\frac{|\Omega|}{\sqrt{\Omega^2\!-\!1}}\,{\cal Z}^<_3[I_1]
    \right],
\end{equation}
\begin{eqnarray}
\label{y11i}
Y_{11}(\omega,\bs{q}) = \frac{1}{\pi}
\left[
     \theta(|\Omega|\leqslant1)\sqrt{1\!-\!\Omega^2}\,
    {\cal Z}_1^>[I_1]
    +
    \theta(|\Omega|\geqslant1)\sqrt{\Omega^2\!-\!1}\,
    {\cal Z}_1^<[I_1]
    \right],
\end{eqnarray}
\begin{equation}
  \label{y12i}
  Y_{12}(\omega,\bs{q}) =- \frac{1}{\pi}
    \theta(|\Omega|\leqslant1)\sqrt{1\!-\!\Omega^2}\,
    {\cal Z}_1^>[I],
\end{equation}
\begin{equation}
  \label{y22i}
  Y_{22}(\omega,\bs{q}) = \frac{1}{\pi}
  \left[
    \theta(|\Omega|\leqslant1)\sqrt{1\!-\!\Omega^2}\,
    {\cal Z}_1^>[I_1]
    +
    \frac{\theta(|\Omega|\geqslant1)}{\sqrt{\Omega^2\!-\!1}}\,{\cal Z}^<_3[I_1]
\right],
\end{equation}
\end{subequations}
where
\begin{subequations}
\label{Zints}
\begin{equation}
\label{z0}
{\cal Z}^>_0[I] = \int\limits_1^\infty dz\sqrt{z^2\!-\!1} \, I(z),
\qquad
{\cal Z}^<_0[I] = \int\limits_0^1 dz\sqrt{1\!-\!z^2} \, I(z),
\end{equation}
\begin{equation}
\label{z1}
{\cal Z}^>_1[I] = \int\limits_1^\infty dz \frac{\sqrt{z^2\!-\!1}}{z^2\!-\!\Omega^2} \, I(z),
\qquad
{\cal Z}^<_1[I] = \int\limits_0^1 dz \frac{\sqrt{1\!-\!z^2}}{\Omega^2\!-\!z^2}\, I(z),
\end{equation}
\begin{equation}
\label{z2}
{\cal Z}^>_2[I] = \int\limits_1^\infty dz \frac{z\sqrt{z^2\!-\!1}}{z^2\!-\!\Omega^2} \, I(z),
\qquad
{\cal Z}^<_2[I] = \int\limits_0^1 dz \frac{z\sqrt{1\!-\!z^2}}{\Omega^2\!-\!z^2}\, I(z),
\end{equation}
\begin{equation}
\label{z3}
{\cal Z}^>_3[I] = \int\limits_1^\infty dz \frac{\left(z^2\!-\!1\right)^{3/2}}{z^2\!-\!\Omega^2} \, I(z),
\qquad
{\cal Z}^<_3[I] = \int\limits_0^1 dz \frac{\left(1\!-\!z^2\right)^{3/2}}{\Omega^2\!-\!z^2}\, I(z).
\end{equation}
\end{subequations}
The functions $I$ and $I_1$ are given by 
\begin{subequations}
\label{izs}
\begin{equation}
\label{i_10}
I_1(z) = \tanh\frac{zQ+W+x}{2}+\tanh\frac{zQ+W-x}{2}-\tanh\frac{zQ-W+x}{2}-\tanh\frac{zQ-W-x}{2},
\end{equation}
\begin{equation}
\label{iz0}
I(z) = \tanh\frac{zQ+W+x}{2}-\tanh\frac{zQ+W-x}{2}-\tanh\frac{zQ-W+x}{2}+\tanh\frac{zQ-W-x}{2}.
\end{equation}
\end{subequations}

\clearpage

\end{widetext}

The scattering rates (\ref{taus_c}) can be expressed in terms of dimensionless integrals as
\begin{equation}
\label{tauij_z_dl}
\tau_{ij}^{-1} = \frac{\alpha_g^2NT}{16\pi^2} 
\left[\frac{NT}{v_g^2\partial n/\partial\mu}\right]
{t}_{ij}^{-1},
\end{equation}
which form the elements of the matrix $\widehat{\textswab{T}}$, see
Eqs.~(\ref{eq}).

In the above integrals the frequency and momentum are expressed in
terms of the dimensionless variables
\begin{equation}
\label{dv}
\bs{Q} = \frac{v_g\bs{q}}{2T},
\qquad
W = \frac{\omega}{2T}.
\end{equation}
Finally, the Coulomb interaction has the form
\begin{equation}
\label{coulomb}
U(\omega,\bs{q})= \frac{2\pi e^2}{q} \widetilde{U} = \frac{2\pi\alpha_g v_g}{q}\widetilde{U}, 
\qquad
\alpha_g=\frac{e^2}{v_g\varepsilon},
\end{equation}
where $\varepsilon$ is the dielectric constant of the environment and
the dimensionless factor $\widetilde{U}$ accounts for screening.

\section{Collision integral in the degenerate regime}
\label{apptaufl}

Here I evaluate the electron-electron scattering rates (\ref{taus_c})
in the degenerate (or Fermi-liquid) regime.

Consider first the functions (\ref{izs}). In the degenerate regime,
$\mu\gg T$, but the frequency $\omega$ is of order $T$, i.e.,
$W\sim1$. As a result, $W\ll x$, and one can expand the functions
(\ref{izs}) in $W$. The expansion is simplified by the following
observation. For $x\gg1$, the difference between two hyperbolic
tangents is sharply peaked at $zQ\sim x$. For positive $z$ one finds
\begin{subequations}
\label{ipfl}
\begin{equation}
I_1(z; x\gg1)\approx - I(z; x\gg1) \approx  
I_P(z; x\gg1),
\end{equation}
\begin{equation}
I_P(z) = \tanh\frac{zQ+W-x}{2}-\tanh\frac{zQ-W-x}{2}.
\end{equation}
\end{subequations}
The relaxation rates (\ref{taus_c}) comprise two integrals each, one
over large and another over small values of the integration variable
$z$, $z\geqslant1$ and $0\geqslant z\geqslant1$, respectively. For
small $z$, the peak at $zQ\sim x$ translates into very large values of
the momentum, $Q>x$. At the same time, typical values of the frequency
are of order temperature, $W\sim1$, meaning $W\ll x$ and hence
${|\Omega|=|W|/Q\ll1}$. This conclusion has the following two
consequences.

First, only the region ${|\Omega|<1}$ contributes to the scattering
rates (up to exponentially small corrections, see below), hence one
only needs to evaluate the integrals (\ref{Zints}) over $z>1$. Taking
into account Eq.~(\ref{ipfl}), I find that all three relaxation rates
(\ref{taus_c}) coincide,
\begin{equation}
\label{mtau}
\tau_{11}=\tau_{12}=\tau_{22},
\end{equation}
such that the matrix of the relaxation rates is degenerate even in the
$2\times2$ sector. This degeneracy is due to the fact that in this
regime only one band contributes to any physical quantity.

Second, the calculation of the single remaining rate,
$\tau_{11}^{-1}$, can be simplified by expanding the function $I_P$ in
powers of $W$,
\begin{equation}
\label{ipw3}
I_P(z) \approx \frac{2W}{1\!+\!\cosh(zQ\!-\!x)}
+ \frac{W^3}{3}\frac{\cosh(zQ\!-\!x)\!-\!2}{[1\!+\!\cosh(zQ\!-\!x)]^2}.
\end{equation}

Noticing that the integrand in Eq.~(\ref{tau_vv}) is an even function
of $W$ and at the same time is independent of the direction of
$\bs{Q}$, the integral can be simplified as
\begin{widetext}
\begin{equation}
\label{tau11fl}
\tau_{11}^{-1} = \alpha_g^2\frac{N}{4\pi}\frac{T^2}{\mu}
\int\limits_0^\infty \frac{dW}{\sinh^2W}
\int\limits_W^\infty QdQ
|\widetilde{U}|^2
\left[
{\cal Z}^>_0[I_P]{\cal Z}^>_1[I_P] - \left(1\!-\!\Omega^2\right)\left({\cal Z}^>_2[I_P]\right)^2
\right].
\end{equation}
The function (\ref{ipfl}) depends only on the combination of
variables, ${y=zQ\!-\!x}$. Changing the integration variable to $y$, I
find for the three functions ${\cal Z}^>_i$ appearing in
Eq.~(\ref{tau11fl})
\begin{subequations}
\label{zys}
\begin{equation}
\label{z0y}
{\cal Z}^>_0[I_P] = \frac{1}{Q^2}\!\int\limits_{Q-x}^\infty\! dy I_P(y,W)\sqrt{(x\!+\!y)^2\!-\!Q^2}
\approx
\frac{1}{Q^2} 
\left[
x J_0 \sqrt{1\!-\!\frac{Q^2}{x^2}} + \frac{J_1}{\sqrt{1\!-\!\frac{Q^2}{x^2}}}
- \frac{J_2Q^2}{2x^3\left(1\!-\!\frac{Q^2}{x^2}\right)^{3/2}}
\right],
\end{equation}
\begin{eqnarray}
&&
{\cal Z}^>_1[I_P] = \!\int\limits_{Q-x}^\infty\! dy I_P(y,W)
\frac{\sqrt{(x\!+\!y)^2\!-\!Q^2}}{(x\!+\!y)^2\!-\!W^2}
\approx
\frac{J_0}{x} \frac{\sqrt{1\!-\!\frac{Q^2}{x^2}}}{1\!-\!\frac{W^2}{x^2}}
-
\frac{J_1}{x^2} 
\frac{1\!+\!\frac{W^2}{x^2}\!-\!2\frac{Q^2}{x^2}}
{\sqrt{1\!-\!\frac{Q^2}{x^2}}\left(1\!-\!\frac{W^2}{x^2}\right)^2}
\\
&&
\nonumber\\
&&
\qquad\qquad\qquad\qquad\qquad\qquad\qquad\qquad\qquad\qquad\qquad\qquad
+
\frac{J_2}{x^3} 
\frac{1\!+\!3\frac{W^2}{x^2}\!-\!\frac{Q^2}{2x^2}\left(3\!+\!\frac{W^2}{x^2}\right)^2
\!+\!\frac{Q^4}{x^4}\left(3\!+\!\frac{W^2}{x^2}\right)}
{\left(1\!-\!\frac{Q^2}{x^2}\right)^{3/2}\left(1\!-\!\frac{W^2}{x^2}\right)^3},
\nonumber
\end{eqnarray}
\begin{eqnarray}
&&
{\cal Z}^>_2[I_P] = \frac{1}{Q}\!\int\limits_{Q-x}^\infty\! dy I_P(y,W)
\frac{(x\!+\!y)\sqrt{(x\!+\!y)^2\!-\!Q^2}}{(x\!+\!y)^2\!-\!W^2}
\approx
J_0 \frac{\sqrt{1\!-\!\frac{Q^2}{x^2}}}{1\!-\!\frac{W^2}{x^2}}
+
\frac{J_1}{x} 
\frac{\frac{Q^2}{x^2}\left(1\!+\!\frac{W^2}{x^2}\right)\!-\!2\frac{W^2}{x^2}}
{\sqrt{1\!-\!\frac{Q^2}{x^2}}\left(1\!-\!\frac{W^2}{x^2}\right)^2}
\\
&&
\nonumber\\
&&
\qquad\qquad\qquad\qquad\qquad\qquad\qquad\qquad\qquad\qquad\qquad
-
\frac{J_2}{x^2} 
\frac{\frac{Q^2}{2x^2}\left(1\!+\!3\frac{W^2}{x^2}\right)\left(3\!+\!\frac{W^2}{x^2}\right)
\!-\!\frac{W^2}{x^2}\left(3\!+\!\frac{W^2}{x^2}\right)
\!-\!\frac{Q^4}{x^4}\left(1\!+\!3\frac{W^2}{x^2}\right)}
{\left(1\!-\!\frac{Q^2}{x^2}\right)^{3/2}\left(1\!-\!\frac{W^2}{x^2}\right)^3},
\nonumber
\end{eqnarray}
\end{subequations}
where
\begin{equation}
\label{jn}
J_n = \!\int\limits_{Q-x}^\infty\! dy I_P(y,W) y^n,
\end{equation}
and the expansion of the algebraic functions in Eqs.~(\ref{zys}) is
justified by the fact that the function $I_P(y)$ has a form of a sharp
peak centered at ${y=0}$ [and in fact exponentially decaying beyond
  $y\sim1$ as can be seen from the expansion (\ref{ipw3})], while
$x\gg1$.

Assuming $W\ll x$ in the degenerate limit, I now disregard the factors
$W^2/x^2$ in the algebraic functions in Eqs.~(\ref{zys}) and form the
integrand in Eq.~(\ref{tau11fl}) as (alternatively one can integrate
the algebraic functions over momentum while keeping the frequency
finite and then compute the limit $x\rightarrow\infty$; the results for
both the leading and subleading terms are the same)
\begin{equation}
\label{z0z1}
{\cal Z}^>_0[I_P]{\cal Z}^>_1[I_P] - (1\!-\!\Omega^2) \left({\cal Z}^>_2[I_P]\right)^2
\approx
\frac{1}{Q^2}\left(1\!-\!\frac{Q^2}{x^2}\right)
\left[
  J_0^2 \frac{W^2}{Q^2} + \frac{J_0J_2\!-\!J_1^2}{x^2} - \frac{J_0J_2}{x^2} \frac{Q^2}{x^2}
  \frac{\frac{11}{2}\!-\!4\frac{Q^2}{x^2}}{\left(1\!-\!\frac{Q^2}{x^2}\right)^2} + \dots
\right],
\end{equation}
where the frequency in the first term comes from the factor
$1\!-\!\Omega^2$ in Eq.~(\ref{tau11fl}), which cannot be neglected
before one determines the order of magnitude of the typical values of
$Q$ in the integral in Eq.~(\ref{tau11fl}).

Consider now the functions $J_n$. The integral for ${n=0}$ yields [the
exact result is followed by an approximation obtained by integrating
Eq.~(\ref{ipw3}); the same approximation can be obtained by series
expansion]
\begin{subequations}
\begin{equation}
\label{jn0}
J_0=2\left[W-\ln\left(e^{Q+W}\!+\!e^x\right)+\ln\left(e^Q\!+\!e^{x+W}\right)\right]
\approx
2W\left(1\!+\!\tanh\frac{x\!-\!Q}{2}\right)-
\frac{4W^3}{3}\frac{\sinh^4\frac{x\!-\!Q}{2}}{\sinh^3(x\!-\!Q)}
\approx
4W \theta(Q<x),
\end{equation}
where the second equality can be directly obtained by using the first
term in the expansion (\ref{ipw3}).

The other functions $J_n$ cannot be integrated in a closed form. One
can either evaluate them approximately using the leading order
expansion (\ref{ipw3}) or compute them numerically. The approximate
calculation yields
\begin{equation}
J_0^2\approx 16 W^2 \theta(Q<x),
\qquad\qquad
J_0J_2\!-\!J_1^2 \approx J_0J_2 \approx \frac{16(\pi^2\!+\!1)}{3}W^2\theta(Q\!+\!2<x).
\end{equation}
\end{subequations}

Substituting the first term in Eq.~(\ref{z0z1}) into Eq.~(\ref{tau11fl}) I find
\[
t_1 = \alpha_g^2\frac{N}{4\pi}\frac{T^2}{\mu}
\int\limits_0^\infty \frac{dW}{\sinh^2W}
\int\limits_W^\infty QdQ
|\widetilde{U}|^2
\frac{W^2}{Q^4}\left(1\!-\!\frac{Q^2}{x^2}\right)
  J_0^2
\approx
\alpha_g^2\frac{4N}{\pi}\frac{T^2}{\mu}
\int\limits_0^\infty \frac{dW W^4}{\sinh^2W}
\int\limits_W^x \frac{dQ}{Q^3}\left(1\!-\!\frac{Q^2}{x^2}\right) |\widetilde{U}|^2.
\]
For bare Coulomb interaction, ${|\widetilde{U}|^2=1}$, I evaluate the
integrals as
\[
\int\limits_0^\infty \frac{dW W^4}{\sinh^2W}
\int\limits_W^x \frac{dQ}{Q^3}\left(1\!-\!\frac{Q^2}{x^2}\right)
=
\frac{1}{2}\int\limits_0^\infty \frac{dW W^4}{\sinh^2W}
\left(\frac{1}{W^2} - \frac{1}{x^2} - \frac{2}{x^2} \ln\frac{x}{W}\right)
=
\frac{\pi^2}{12} - \frac{\pi^4}{60x^2}\left(1+2\ln\frac{\gamma\mu}{T}\right),
\qquad
\gamma=0.474.
\]
For statically screened Coulomb,
${|\widetilde{U}|^2=Q^2/(Q+\varkappa)^2}$, where the Thomas-Fermi
screening yields ${\varkappa=N\alpha_gx}$, I find
\[
\int\limits_W^x \frac{dQ}{Q}\frac{\left(1\!-\!\frac{Q^2}{x^2}\right)}{(Q+\varkappa)^2}
=
\left[\frac{1}{\varkappa^2}-\frac{1}{x^2}\right]\!\!
\left[\frac{W}{W+\varkappa}-\frac{x}{x+\varkappa}\right]
+
\left[\frac{1}{\varkappa^2}+\frac{1}{x^2}\right]\!\!
\left[\ln\frac{x}{x+\varkappa}-\ln\frac{W}{W+\varkappa}\right]
+
\frac{1}{x^2}\ln\frac{W}{x},
\]
which recovers the above result in the limit
$\varkappa\rightarrow0$. The limit $x\rightarrow\infty$ here is
non-trivial since $\varkappa\propto x$ and one can discern three
regimes: $\varkappa\ll 1$, $1\ll\varkappa\ll x$, and $\varkappa\gg
x$. In the first regime, the integral differs from the above result
for $\varkappa=0$ by small corrections. The last regime has only a
formal interest, since in order to achieve this one has to consider
large $\alpha_g$ contradicting the assumptions of the present
approach. In the second regime, the above expression can be
represented as a series expansion in $x^{-1}$. For not too small
$\alpha_g\gtrsim0.1$
\[
\int\limits_W^x \frac{dQ}{Q}\frac{\left(1\!-\!\frac{Q^2}{x^2}\right)}{(Q+\varkappa)^2}
=
\frac{1}{N^2\alpha_g^2x^2}\left[\ln\frac{N\alpha_g}{W}-g(\alpha_g)\right]
+
\frac{W}{2N^2\alpha_g^3x^3}
+{\cal O}(x^{-4}),
\]
where
\[
g(\alpha_g)=1-N\alpha_g+\ln(1\!+\!N\alpha_g)
+N^2\alpha_g^2\ln\frac{1\!+\!N\alpha_g}{N\alpha_g},
\qquad
g(\alpha_g\gtrsim0.25)
\approx
\ln(1\!+\!N\alpha_g)+\frac{1}{2}.
\]
Note that the subleading term in this expansion is essential since the
leading term changes sign at not too large $W$, while the integral is
explicitly positive. Integrating over the frequencies, I find
\[
\int\limits_0^\infty \frac{dW W^4}{\sinh^2W}
\int\limits_W^x \frac{dQ}{Q}\frac{\left(1\!-\!\frac{Q^2}{x^2}\right)}{(Q+\varkappa)^2}
=
\frac{\pi^4}{30N^2\alpha_g^2x^2}\left[\ln(\gamma N\alpha_g)-g(\alpha_g)\right]
+
\frac{15\zeta(5)}{4N^2\alpha_g^3x^3}.
\]
Again, while formally subleading in the limit $x\rightarrow\infty$,
the second term is important since for small $\alpha_g$ the first term
is negative.

Consider now the contribution of the second term in
Eq.~(\ref{z0z1}). Similarly to the above, I find
\[
t_2 = \alpha_g^2\frac{N}{4\pi}\frac{T^2}{\mu}
\int\limits_0^\infty \frac{dW}{\sinh^2W}
\int\limits_W^\infty QdQ
\frac{1}{Q^2}\left(1\!-\!\frac{Q^2}{x^2}\right)
\frac{J_0J_2-J_1^2}{x^2}
\approx
\alpha_g^2\frac{4N}{3\pi}(\pi^2\!+\!1)\frac{T^2}{\mu x^2}
\int\limits_0^\infty \frac{dW W^4}{\sinh^2W}
\int\limits_W^{x-2} \frac{dQ}{Q}\left(1\!-\!\frac{Q^2}{x^2}\right),
\]
which in the limit $x\rightarrow\infty$ is clearly subleading to $t_1$.

\end{widetext}

As a result, the ``scattering rate'' $\tau_{11}^{-1}$ in the
degenerate regime is given by
\begin{subequations}
\label{taueefl}
\begin{equation}
\label{ttreefl}
\tau_{11}^{-1} \approx 
\frac{\pi N}{3} \alpha_g^2 \frac{T^2}{\mu}.
\end{equation}
for the unscreened Coulomb interaction. Taking into account screening,
the result in the most interesting regime, $1\ll\varkappa < x$ is
\begin{equation}
\label{ttreefltf}
\tau_{11}^{-1} \approx 
\frac{T^4}{\mu^3}\left[\frac{\pi}{30}\left[\ln(\gamma N\alpha_g)-g(\alpha_g)\right]
+
\frac{15\zeta(5)}{4\pi\alpha_gx}\right].
\end{equation}
\end{subequations}
One should keep in mind, however, that the Thomas-Fermi screening is
nothing but the static limit of the random phase approximation
(RPA). The latter is by no means exact, especially for not so small
coupling constants. Hence including the RPA screening, but neglecting
vertex corrections might not yield the better approximation to the
exact result than the bare Coulomb interaction. At the neutrality
point this is supported by $1/N$ expansion \cite{mfss,schutt} where
one can show that for the momentum range dominating the integration
the leading-order results are given by the unscreened Coulomb
(perhaps, with the renormalized coupling constant). While no such
analysis has been reported in the degenerate regime, the above
statement seems plausible on physical grounds and hence
Eq.~(\ref{ttreefltf}) should be treated with care.

\section{Collision integral close to charge neutrality}
\label{apptaudp}

At charge neutrality ${I(x\!=\!0)=0}$ and hence
${\tau_{12}^{-1}=0}$. In vicinity of charge neutrality the dependence
on the chemical potential (or on the dimensionless variable
${x=\mu/T}$) comes from the compressibility in the dimensionfull
prefactor in Eq.~(\ref{tauij_z_dl}) and the functions $I$ and $I_1$
determining the dimensionless scattering rates $t_{ij}^{-1}$.

The prefactor in Eq.~(\ref{tauij_z_dl}) can be expressed as follows
\[
\frac{\alpha_g^2NT}{16\pi^2} 
\left[\frac{NT}{v_g^2\partial n/\partial\mu}\right]
\approx
\frac{\alpha_g^2T}{4\pi\ln2}\left(1-\frac{x^2}{8\ln2}\right).
\]
Expanding both functions (\ref{izs}) for
${x\ll1}$, one finds
\begin{subequations}
\label{izsdp}
\begin{eqnarray}
\label{i_10dp}
&&
I_1 = I_1^{(0)} + x^2 I_1^{(2 )}+ {\cal O}(x^3),
\\
&&
\nonumber\\
&&
I_1^{(0)} = \frac{4 \sinh W}{\cosh W \!+\! \cosh zQ},
\nonumber\\
&&
\nonumber\\
&&
I_1^{(2 )} = \frac{2\sinh W (\cosh 2zQ\!-\!2\cosh W\cosh zQ\!-\!3)}{(\cosh W\!+ \!\cosh zQ)^3}
\nonumber
\end{eqnarray}
\begin{equation}
\label{iz0dp}
I = -xI^{(1)}\!+ {\cal O}(x^3),
\quad
I^{(1)}\! = 4 \frac{\sinh zQ \sinh W}{(\cosh W \!+\! \cosh zQ)^2}.
\end{equation}
\end{subequations}

Substituting these expansions into Eqs.~(\ref{taus_c}), one can
establish the leading terms in the expansion of the scattering rates
\begin{subequations}
\label{tausdp}
\begin{equation}
\label{tau11dp}
\frac{1}{\tau_{11}} = \frac{\alpha_g^2T}{4\pi\ln2}
\left[\frac{1}{t_{11}^{(0)}} 
+ x^2\!\left(\frac{1}{t_{11}^{(2)}}\!-\!\frac{1}{8\ln2}\frac{1}{t_{11}^{(0)}}\right)\! 
+ {\cal O}(x^3)\right]\!,
\end{equation}
\begin{equation}
\label{tau12dp}
\frac{1}{\tau_{12}} = \frac{\alpha_g^2T}{4\pi\ln2}\frac{x}{t_{12}^{(1)}} + {\cal O}(x^3),
\end{equation}
\begin{equation}
\label{tau22dp}
\frac{1}{\tau_{22}} = \frac{\alpha_g^2T}{4\pi\ln2}
\left[\frac{1}{t_{22}^{(0)}} 
+ x^2\!\left(\frac{1}{t_{22}^{(2)}}\!-\!\frac{1}{8\ln2}\frac{1}{t_{22}^{(0)}}\right)\! 
+ {\cal O}(x^3)\right]\!.
\end{equation}
\end{subequations}
The quantities $t_{ij}^{(0,1,2)}$ in Eqs.~(\ref{tausdp}) are the
expansion coefficients of the dimensionless integral in
Eq.~(\ref{tauij_z_dl}) close to the Dirac point in the self-evident
notation similar to that in Eqs.~(\ref{izsdp}). For unscreened Coulomb
interaction these quantities are just numbers without any dependence
on any physical parameter. If screening is taken into account, then
these numbers depend on the screening length, i.e. on a fixed
combination of the coupling constant and temperature.

For unscreened Coulomb interaction, one finds the following numerical
values (neglecting the small \cite{kash} exchange contribution):
\[
\left(t_{11}^{(0)}\right)^{-1} \approx 33.13,
\quad
\left(t_{11}^{(2)}\right)^{-1} \approx 3.38,
\]
\[
\left(t_{12}^{(1)}\right)^{-1} \approx 5.45,
\quad
\left(t_{22}^{(0)}\right)^{-1} \approx 18.02,
\quad
\left(t_{22}^{(2)}\right)^{-1} \approx 4.73.
\]

\section{Optical conductivity close to charge neutrality}
\label{appsomdp}

Inverting the matrix $\textswab{S}_{xx}$ using the identity
\[
\left[\textswab{S}_{xx}(0) + \delta \textswab{S}_{xx}\right]^{-1}
\approx
\textswab{S}_{xx}^{-1}(0)-
\textswab{S}_{xx}^{-1}(0)\delta \textswab{S}_{xx}\textswab{S}_{xx}^{-1}(0),
\]
and substituting the result into the general expression (\ref{sigma0})
together with the expansions of the matrices
$\widehat{\textswab{M}}_{h(n)}$, one finds the leading contribution to
Eq.~(\ref{som0}) as
\begin{eqnarray*}
&&
\widehat{\textswab{M}}_{h}
\textswab{S}_{xx}^{-1}
\widehat{\textswab{M}}_{n}
\approx
\widehat{\textswab{M}}_{h}
\textswab{S}_{xx}^{-1}(0)
\widehat{\textswab{M}}_{n}
+
\\
&&
\\
&&
\qquad
+
\widehat{\textswab{M}}_{h}(0)
\textswab{S}_{xx}^{-1}(0)\delta \textswab{S}_{xx}\textswab{S}_{xx}^{-1}(0)
\widehat{\textswab{M}}_{n}(0)
\\
&&
\\
&&
\qquad
+
\delta\widehat{\textswab{M}}_{h}
\textswab{S}_{xx}^{-1}(0)\delta \textswab{S}_{xx}\textswab{S}_{xx}^{-1}(0)
\widehat{\textswab{M}}_{n}(0)
\\
&&
\\
&&
\qquad
+
\widehat{\textswab{M}}_{h}(0)
\textswab{S}_{xx}^{-1}(0)\delta \textswab{S}_{xx}\textswab{S}_{xx}^{-1}(0)
\delta\widehat{\textswab{M}}_{n}.
\end{eqnarray*}
The first line comprises the zeroth order result, Eq.~(\ref{som0}),
and the correction
\[
\delta\sigma_1(\omega)=
\frac{\gamma_{11}e^2Tx^2}{-i\omega\!+\!\tau_{\rm dis}^{-1}\!+\!\tau_{11}^{-1}(0)}
+
\frac{\gamma_{12}e^2Tx^2}{-i\omega\!+\!\tau_{\rm dis}^{-1}\!+\!\gamma_{13}\tau_{22}^{-1}(0)},
\]
where the scattering rates are evaluated at the Dirac point, ${x=0}$, and
the numerical coefficients are
\[
\gamma_{11}=\frac{2\ln2}{\pi}\frac{8\ln2}{27\zeta(3)}\approx0.17\times\frac{2\ln2}{\pi}
\approx0.075,
\]
\[
\gamma_{12}=\frac{3[4\pi^2\ln2-27\zeta(3)]^2}{162\pi\zeta(3)\ln2-\pi^5}
\approx0.66,
\]
\[
\gamma_{13}=\frac{162\pi\zeta(3)\ln2}{162\pi\zeta(3)\ln2-\pi^5}
\approx3.59.
\]
The second line yields
\[
\delta\sigma_2(\omega)=
\frac{e^2Tx^2}{2\pi^2}
\frac{\frac{1}{\tau_{11}^{(2)}}-\frac{1}{8\ln2}\frac{1}{\tau_{11}(0)}}
     {\left[-i\omega\!+\!\tau_{\rm dis}^{-1}\!+\!\tau_{11}^{-1}(0)\right]^2},
\]
where the numerator represents the leading correction to the
scattering rate $\tau_{11}^{-1}$, see Eq.~(\ref{tau11dp}). Finally,
the last two lines give rise to two similar contributions
\begin{eqnarray*}
&&
\delta\sigma_3(\omega)=
\frac{e^2Tx^2}{-i\omega\!+\!\tau_{\rm dis}^{-1}\!+\!\tau_{11}^{-1}(0)}
\frac{1}{-i\omega\!+\!\tau_{\rm dis}^{-1}\!+\!\gamma_{13}\tau_{22}^{-1}(0)}
\\
&&
\\
&&
\qquad\qquad
\times
\left[
\gamma_{31} \tau_{22}^{-1}(0)
+
\gamma_{32} \left(-i\omega\!+\!\tau_{\rm dis}^{-1}\right)
-
\gamma_{33}/\tau_{12}^{(1)}
\right],
\end{eqnarray*}
\begin{eqnarray*}
&&
\delta\sigma_4(\omega)=
\frac{e^2Tx^2}{-i\omega\!+\!\tau_{\rm dis}^{-1}\!+\!\tau_{11}^{-1}(0)}
\frac{1}{-i\omega\!+\!\tau_{\rm dis}^{-1}\!+\!\gamma_{13}\tau_{22}^{-1}(0)}
\\
&&
\\
&&
\qquad\qquad
\times
\left[
\gamma_{41} \left(-i\omega\!+\!\tau_{\rm dis}^{-1}\right)
+
\gamma_{42}/\tau_{12}^{(1)}
\right],
\end{eqnarray*}
where
\[
\gamma_{31}=\frac{288\ln^32}{162\pi\zeta(3)\ln2-\pi^5}
\approx0.81,
\]
\[
\gamma_{32}=\frac{3(96\ln^32+27\zeta(3)-8\pi^2\ln2)}{162\pi\zeta(3)\ln2-\pi^5}
\approx0.246,
\]
\[
\gamma_{33}=\frac{3[8\pi^2\ln2-27\zeta(3)]}{162\pi\zeta(3)\ln2-\pi^5}
\approx0.57,
\]
\[
\gamma_{41}=\frac{3[4\pi^2\ln2-27\zeta(3)][8\pi^2\ln2-54\zeta(3)]}{2(162\pi\zeta(3)\ln2-\pi^5)}
\approx0.66,
\]
\[
\gamma_{42}=\frac{81\zeta(3)[4\pi^2\ln2-27\zeta(3)]}{2\pi(162\pi\zeta(3)\ln2-\pi^5)}
\approx0.67.
\]

\clearpage

\bibliography{viscosity_refs}

\end{document}